\documentclass[12pt]{article}
\usepackage{jheppub}

\pdfoutput=1

\usepackage{amsmath,bbm,array,amsfonts,graphicx,wrapfig,lscape,float,mathtools,multirow,longtable}
\usepackage{physics}
\usepackage{subcaption}
\usepackage{braket}
\usepackage{cancel}
\usepackage{tikz}
\usetikzlibrary{decorations.pathreplacing,decorations.markings,arrows.meta}
\definecolor{yellow2}{rgb}{0.98, 0.80, 0.20}
\definecolor{blue2}{RGB}{65,128,255}
\usepackage[colorlinks=true]{hyperref}

\renewcommand{\vev}[1]{\langle #1 \rangle}
\newcommand{\floor}[1]{ \left\lfloor{ #1} \right\rfloor}
\renewcommand{\d}{\partial }

\definecolor{yellow2}{rgb}{0.98, 0.80, 0.20}
\definecolor{darkspringgreen}{rgb}{0.09, 0.45, 0.27}
\definecolor{forestgreen}{rgb}{0.13, 0.55, 0.13}

\newcommand{\be}{\begin{equation}}
\newcommand{\ee}{\end{equation}}
\newcommand{\beq}{\begin{equation}}
\newcommand{\beql}[1]{\begin{equation}\label{#1}}
\newcommand{\eeq}{\end{equation}}
\newcommand{\ba}{\begin{array}}
\newcommand{\ea}{\end{array}}
\newcommand{\bea}{\begin{eqnarray}}
\newcommand{\beal}[1]{\begin{eqnarray}\label{#1}}
\newcommand{\eea}{\end{eqnarray}}
\newcommand{\ben}{\begin{enumerate}}
\newcommand{\een}{\end{enumerate}}
\newcommand{\bean}{\begin{eqnarray*}}
\newcommand{\eean}{\end{eqnarray*}}
\newcommand{\eref}[1]{(\ref{#1})}
\newcommand{\sref}[1]{\S\ref{#1}}

\newcommand{\fref}[1]{Figure \ref{#1}}
\newcommand{\btab}[1]{\begin{tabular}{#1}}
\newcommand{\etab}{\end{tabular}}

\def \Black {\pdfsetcolor{0 0 0 1}}

\newcommand{\comment}[1]{}

\newcommand{\CN}{{\cal N}}

\newcommand{\ud}{\mathrm{d}}

\newcommand{\qed}{\nobreak \ifvmode \relax \else
      \ifdim\lastskip<1.5em \hskip-\lastskip
      \hskip1.5em plus0em minus0.5em \fi \nobreak
      \vrule height0.75em width0.5em depth0.25em\fi}

\title{Graded Quivers, Generalized Dimer Models and Toric Geometry}

\author[a,b,c]{Sebasti\'an Franco,} 
\author[a,b]{Azeem Hasan}

\affiliation[a]{
Physics Department, The City College of the CUNY \\
160 Convent Avenue, New York, NY 10031, USA}

\affiliation[b]{Physics Program and $^c$Initiative for the Theoretical Sciences \\
The Graduate School and University Center, The City University of New York  \\
365 Fifth Avenue, New York NY 10016, USA}

\emailAdd{sfranco@ccny.cuny.edu}
\emailAdd{ahasan@gradcenter.cuny.edu}

\abstract{The open string sector of the topological B-model model on CY $(m+2)$-folds is described by $m$-graded quivers with superpotentials. This correspondence extends to general $m$ the well known connection between CY $(m+2)$-folds and gauge theories on the worldvolume of D$(5-2m)$-branes for $m=0,\ldots, 3$. We introduce $m$-dimers, which fully encode the $m$-graded quivers and their superpotentials, in the case in which the CY $(m+2)$-folds are toric. Generalizing the well known $m=1,2$ cases, $m$-dimers significantly simplify the connection between geometry and $m$-graded quivers. A key result of this paper is the generalization of the concept of perfect matching, which plays a central role in this map, to arbitrary $m$. We also introduce a simplified algorithm for the computation of perfect matchings, which generalizes the Kasteleyn matrix approach to any $m$. We illustrate these new tools with a few infinite families of CY singularities.
}

\begin{document}
    \maketitle

\section{Introduction}

D-branes probing singularities provide a powerful framework for engineering quantum field theories in various dimensions and studying their dynamics. In particular, a large class of $(6-2m)$-dimensional gauge theories can be realized in Type IIB string theory on the worldvolume of D$(5-2m)$-branes probing Calabi-Yau (CY) $(m + 2)$-folds. The prototypical example involves D3-branes on CY 3-folds \cite{Morrison:1998cs,Beasley:1999uz,Feng:2000mi,Beasley:2001zp,Feng:2001xr,Feng:2001bn,Feng:2002zw,Wijnholt:2002qz,Benvenuti:2004dy,Franco:2005rj,Benvenuti:2005ja,Franco:2005sm,Butti:2005sw}. While in the context of quantum field theories and string theory we are restricted to the $0\leq m\leq 3$ range, these constructions can be extended to arbitrary $m\geq 0$ in the framework of $m$-graded quivers \cite{Franco:2017lpa}, building on the mathematical ideas in \cite{Ginzburg:2006fu, MR3590528,MR2553375}.\footnote{In what follows, we will often use the terms graded quiver and gauge theory interchangeably.} The physical relevance of $m$-graded quivers for general $m$ is that they describe the open string sector of the topological B-model on CY $(m + 2)$-folds \cite{Aspinwall:2008jk,lam2014calabi,Closset:2018axq}. 

Given an $m$-graded quiver arising from such a B-model setup, it is natural to ask what the corresponding CY$_{m+2}$ is. Similarly, starting from a CY$_{m+2}$ it is interesting to determine the corresponding quiver theory.\footnote{In fact, the correspondence between CY$_{m+2}$'s and quiver theories is not one-to-one but one-to-many due to dualities and their generalizations.} There are multiple approaches for addressing these questions which can become, in practice, computationally challenging.

The dictionary connecting geometry and quivers is particularly well understood when the CY $(m+2)$-folds are toric, in terms of objects that generalize dimer models. In this case, T-duality connects the D$(5-2m)$-branes probing CY $(m+2)$-folds to new configurations of branes living on $\mathbb{T}^{m+1}$. For $m=1$, 2 and 3, these configurations are known as brane tilings \cite{Franco:2005rj,Franco:2005sm}, brane brick models \cite{Franco:2015tya,Franco:2016nwv,Franco:2016qxh} and brane hyperbrick models \cite{Franco:2016tcm}, respectively. In this paper we will generalize these constructions, developing similar objects that have been postulated to describe the graded quivers associated to toric CY $(m+2)$-folds for general values of $m$ \cite{Futaki:2014mpa,Franco:2016qxh,Franco:2017lpa,Closset:2018axq}. We will collectively refer to these combinatorial objects as {\it generalized dimer models}.

Generalized dimer models significantly streamline the connection between graded quivers and the underlying CY's. This problem has been extensively studied for CY 3-folds and 4-folds. For toric CY 3-folds, {\it perfect matchings} are one of the main ingredients in the map between gauge theory and geometry \cite{Franco:2005rj}. For toric CY 4-folds, perfect matchings generalize to {\it brick matchings}, which were introduced in \cite{Franco:2015tya}. The power of perfect matchings and brick matchings follows from the fact that they admit combinatorial definitions in terms of the underlying brane tilings and brane brick models. In this paper we will introduce {\it generalized perfect matchings}, which parameterize the toric CY $(m+2)$-folds associated to $m$-graded quivers. In the standard case of brane tilings, the simplification is even more striking, thanks to the existence of a straightforward algorithmic approach for finding perfect matchings based on the Kasteleyn matrix \cite{2003math.....10326K,2003math.ph..11005K,Hanany:2005ve,Franco:2005rj,Franco:2012mm}. In this paper we will also introduce a generalization of the Kasteleyn matrix procedure to all $m$. 

This paper is organized as follows. In \sref{section_graded_quivers} we review $m$-graded quivers and their mutations. In \sref{section_brane_tilings_and_BBMs} we review brane tilings and brane brick models. In \sref{section_generalized_dimers} we introduce $m$-dimers, which encode the $m$-graded quivers associated to toric CY $(m+2)$-folds. In \sref{section_Cm+2_permutohedra} we discuss the $m$-dimers associated to $\mathbb{C}^{m+2}$, which consist of $(m+2)$-permutohedra. In \sref{section_perfect_matchings} we generalize the concept of perfect matchings to $m$-dimers with arbitrary $m$. We provide a first definition of perfect matchings based on the superpotential. We also explain how to use them to compute the corresponding toric diagram. In \sref{section_F0m} we apply these ideas to determine de moduli space of the $F_0^{(m)}$ infinite family of quiver theories. In \sref{section_simplified_algorithm} we present a second definition of perfect matchings, based on chiral cycles, and a simplified algorithm for their computation based on a Grassmann integral. We apply this method for calculating the moduli space of the $Y^{1,0}(\mathbb{P}^m)$ family in \sref{section_Y10}. In \sref{section_orbifolds_Cm+2} we present a general discussion of abelian orbifolds of $\mathbb{C}^{m+2}$. We present our conclusions in \sref{label_section_conclusions}. Appendices \ref{section_pms_F0} and \ref{section_orbifold_pms_appendix} contain additional details on the perfect matchings for $F_0^{(m)}$ and orbifolds of $\mathbb{C}^{m+2}$.

\section{Graded quivers}
 
 \label{section_graded_quivers}
  
A central aim of this paper is to develop new tools to connect $m$-graded quivers to toric CY $(m+2)$-folds. In order to make our presentation self-contained, in this section we briefly review $m$-graded quivers and their connections to physics. We refer the reader to \cite{Franco:2017lpa,Closset:2018axq} for in-depth presentations and to \cite{lam2014calabi} for a mathematical analysis. Related works include \cite{Closset:2017yte,Closset:2017xsc, Eager:2018oww}.

Given an integer $m \geq 0$, an $m$-graded quiver is a quiver equipped with a grading for every arrow $\Phi_{ij}$ by a {\it quiver degree}:
\beq
|\Phi_{ij}| \in \{ 0, 1, \cdots, m\}~.
\eeq
To every node $i$ we associate a unitary ``gauge group" $U(N_i)$. Arrows stretching between nodes correspond to bifundamental or adjoint ``fields".\footnote{The framework of $m$-graded quivers can be extended to theories with gauge groups that are not unitary and with fields that do not transform in the bifundamental or adjoint representations, i.e. theories that are not of quiver type. We will not consider these possibilities in this paper.}

For every arrow $\Phi_{ij}$, its conjugate has the opposite orientation and degree $m-|\Phi_{ij}|$:
\beq\label{Phi opp intro}
\overline{\Phi}_{ji}^{(m-c)}\equiv \overline{(\Phi_{ij}^{(c)})}~.
\eeq
Here we introduced a notation that will be used throughout the paper, in which the superindex explicitly indicates the degree of the corresponding arrow, namely $|\Phi_{ij}^{(c)}|=c$. 

Since the integer $m$ determines the possible degrees, different values of $m$ give rise to qualitatively different classes of graded quivers. The different types of arrows can be restricted to have degrees in the range:
\beq\label{arrows fields}
\Phi_{ij}^{(c)} \; : i \longrightarrow j~, \qquad c=0, 1, \cdots, n_c-1~, \qquad n_c \equiv    \floor{m+2\over 2}~,
\eeq
since other degrees can be obtained by conjugation.\footnote{The range of degrees in \eref{arrows fields} is just a conventional choice. The $n_c$ ``fundamental" degrees can be picked differently. Moreover, as we will later illustrate in examples, sometimes it is convenient to deal with all possible values of the degrees. For every arrow, either $\Phi_{ij}^{(c)}$ or $\overline{\Phi}_{ji}^{(m-c)}$ can be regarded as the fundamental object, while the other one is its conjugate.} We refer to degree 0 fields as {\it chiral fields}.

Graded quivers for $m=0,1,2,3$ describe $d=6,4,2,0$ minimally supersymmetric gauge theories, respectively. Different degrees map to different types of superfields. The correspondence between graded quivers and gauge theories is summarized in \eref{table_graded_quivers_QFTs}, where we also indicate how some of these theories can be engineered in terms of Type IIB D$(5-2m)$-branes probing CY $(m+2)$-folds.
\be
\begin{tabular}{c|cccc}
$m$ & $0$ &$1$& $2$ & $3$  \\
 \hline
CY    &CY$_2$ &CY$_3$ & CY$_4$& CY$_5$\\
SUSY & 6d  $\CN=(0,1)$ & 4d $\CN=1$ & 2d $\CN=(0,2)$ & 0d $\CN=1$ 
\end{tabular}
\label{table_graded_quivers_QFTs}
\ee

\paragraph{Superpotential.}
Graded quivers admit {\it superpotentials}, which are given by linear combinations of gauge invariant terms of degree $m-1$:
\be
W= W(\Phi)~,\qquad\qquad  |W|= m-1~.
\label{superpotential_degree}
\ee
Gauge invariant terms correspond to closed oriented cycles in the quiver, which might involve conjugation of some of the arrows. The superpotential encodes relations on the path algebra of the form $\partial_\Phi W=0$.

There is no possible superpotential for $m=0$. For $m=1,2,3$, the superpotentials take the schematic forms:
\beq
\begin{array}{ll}
m=1 : \ &W= W(\Phi^{(0)})~, \\[.35cm]
m=2 : \  &W= \Phi^{(1)}J(\Phi^{(0)})+ \overline{\Phi}^{(1)} E(\Phi^{(0)})~, \\[.35cm]
m=3 : \  &W= \Phi^{(1)}\Phi^{(1)} H(\Phi^{(0)})+ \Phi^{(2)} J(\Phi^{(0)})~, 
\end{array}
\label{W_different_m}
\eeq
where $W(\Phi^{(0)})$, $J(\Phi^{(0)}), E(\Phi^{(0)})$ and $H(\Phi^{(0)})$ are holomorphic functions of the chiral fields.

\paragraph{Kontsevich bracket condition.}
In addition to the constraint on its degree \eref{superpotential_degree}, the superpotential must also satisfy:
\beq
\{W,W \}=0 ~.
\label{superpotential_Kontsevich}
\eeq 
Here $\{ f, g \}$ denotes the Kontsevich bracket, which is a natural generalization of the Poisson bracket to a graded quiver and is defined as follows   
\be
\{ f, g \}= \sum_\Phi \left( {\d f\over \d \Phi}{\d g\over \d\overline{\Phi}} +(-1)^{(|f|+1)|\overline{\Phi}|+(|g|+1)|\Phi|+ |\Phi||\overline{\Phi}|+1} {\d f\over \d \overline{\Phi}}{\d g\over \d \Phi} \right)~.
\ee

The degree and Kontsevich bracket constraints on the superpotential are necessary for the good behavior of a differential operator that can be associated to graded quivers. See \cite{Franco:2017lpa,Closset:2018axq} for details.

In \sref{section_generalized_dimers} we will discuss how in the case of graded quivers related to toric CY $(m+2)$-folds the superpotential has additional structure. These extra features are at the heart of their description in terms of generalized dimer models.

\subsection{Mutations}
 
 $m$-graded quivers admit order $m+1$ mutations. For $m\leq 3$ these mutations reproduce the dualities of the corresponding gauge theories, namely: no duality for $6d$ $\mathcal{N}=(0,1)$, Seiberg duality for $4d$ $\mathcal{N}=1$ \cite{Seiberg:1994pq}, triality for $2d$ $\mathcal{N}=(0,2)$ \cite{Gadde:2013lxa} and quadrality for $0d$ $\mathcal{N}=1$ \cite{Franco:2016tcm}. Moreover, the mutations provide a generalization of these dualities to $m>3$. It is natural to expect that these generalized dualities correspond to mutations of exceptional collections of B-branes in CY $(m+2)$-folds \cite{Franco:2017lpa}. 
 
A precise prescription determines the transformation of the quiver and superpotential under mutations. We refer the reader to \cite{Franco:2017lpa,Closset:2018axq} for detailed presentations.

\subsection{Generalized anomaly cancellation}

Under a mutation at a node $\star$, its rank transform according to:
\beq
N'_\star = N_0 - N_\star ~,
\eeq
where $N_0$ is the total number of incoming chiral fields at node $\star$. Demanding invariance of the ranks under $m+1$ consecutive mutations of the same node leads to the {\it generalized anomaly cancellation} conditions. 
For $m$ odd, these conditions are given by:
\beq
\sum_j N_j \sum_{c=0}^{n_c-1} (-1)^c \left(\CN(\Phi_{ji}^{(c)})-\CN(\Phi_{ij}^{(c)})\right)=0~, \qquad \forall i~, \qquad {\rm if}\;\; m \in 2\mathbb{Z}+1~,
\label{anomaly_odd}
\ee
where $\CN(\Phi_{ij}^{(c)})$ denotes the number of arrows from $i$ to $j$ of degree $c$. For every fixed $i$, the sum over $j$ runs over all nodes in the quiver (including $i$), and $n_c$ is given by \eref{arrows fields}. For $m$ even, the conditions become
\be
\sum_j N_j \sum_{c=0}^{n_c-1}(-1)^c \left(\CN(\Phi_{ji}^{(c)})+\CN(\Phi_{ij}^{(c)})\right)=2N_i~, \qquad \forall i~, \qquad {\rm if}\;\; m \in 2\mathbb{Z}~.
\label{anomaly_even}
\ee
For $m=0,1,2,3$, these conditions reproduce the cancellation of non-abelian anomalies for the corresponding $d=6,4,2,0$ gauge theories with gauge group $\prod_i U(N_i)$.

\section{Brane tilings and brane brick models}
 
 \label{section_brane_tilings_and_BBMs}

Before introducing generalized dimer models, we present a brief review of {\it brane tilings} \cite{Franco:2005rj,Franco:2005sm} and {\it brane brick models} \cite{Franco:2015tya,Franco:2016nwv}. These objects have been extensively studied in the literature. The aim of this section is just to highlight some basic properties that we will later generalize to arbitrary $m$.

\subsection{Brane tilings}

\label{section_brane_tilings}

The $4d$ $\mathcal{N} = 1$ gauge theories living on the worldvolume of D3-branes probing toric CY$_3$ singularities are fully encoded by bipartite graphs on $\mathbb{T}^2$ denoted brane tilings \cite{Franco:2005rj,Franco:2005sm}.\footnote{Here and in the discussions that will follow for general $m$, we focus on {\it toric phases} of the quiver theories. Such phases exist for any toric CY and can be defined as theories that are fully captured by generalized dimers or, equivalently, periodic quivers. Starting from them, non-toric phases can be generated by mutations.} In fact a brane tiling is a physical brane configuration consisting of an NS5-brane wrapping a holomorphic curve from which D5-branes are suspended, which is related to the D3-branes probing the CY$_3$ by T-duality.\footnote{We refer to both the full fledged brane configuration and the bipartite graph representing the most important features of its structure as brane tiling. We will adopt a similar approach when discussing generalized dimer models.} The holomorphic surface is given by the zero locus of the Newton polynomial of the toric diagram.

\begin{figure}[t]
	\centering
	\includegraphics[width=12cm]{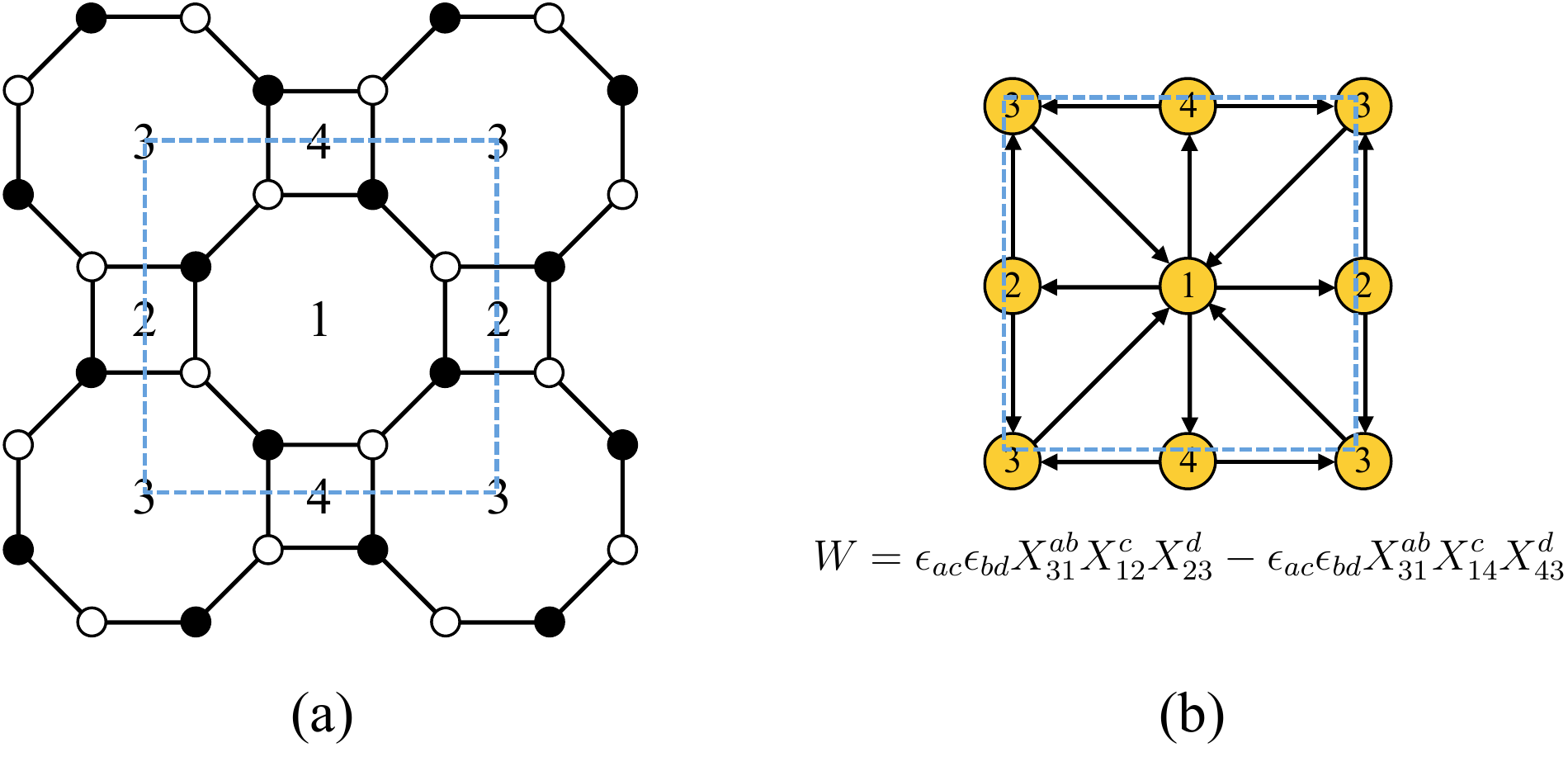}
\caption{a) Brane tiling and b) dual periodic quiver for phase 2 of $F_0$. The unit cell is indicated with dashed blue lines. We also show the superpotential, making the $SU(2)\times SU(2)$ global symmetry of this theory manifest.}
	\label{F0_tiling_quiver}
\end{figure}

A simple dictionary relates brane tilings to the corresponding gauge theories. Faces, edges and nodes in the tiling correspond to unitary gauge group factors, bifundamental or adjoint chiral fields and superpotential terms (with sign determined by the node color), respectively. \fref{F0_tiling_quiver} illustrates these ideas with an explicit example. This theory is often referred to as phase 2 of $F_0$ \cite{Feng:2002zw}. The 4 gauge groups, 12 chiral fields and 8 superpotential terms are easily read from the brane tiling. Equivalently, the same information is captured by a periodic quiver on $\mathbb{T}^2$, which is obtained from the brane tiling by graph dualization. Like a brane tiling, a periodic quiver not only summarizes the matter content and gauge symmetry of a theory but also its superpotential, which is encoded in its minimal plaquettes. For detailed discussions of brane tilings see e.g. \cite{Franco:2005rj,Kennaway:2007tq,Yamazaki:2008bt} and references therein.

The gauge theories associated to toric CY$_3$'s satisfy the so-called {\it toric condition}, namely that every chiral field belongs to exactly two terms in the superpotential with opposite signs. Equivalently, every arrow in the periodic quiver belongs to two adjacent plaquettes with opposite orientations. Brane tilings automatically implement this condition, since every edge connects two nodes of opposite colors.

Due to the toric condition of the superpotential, all the $F$-terms are of the form:
\beq
    \pdv{W}{X_{i}} = M_{i}^{+}(X_{j}) - M_{i}^{-}(X_{j}) ~, 
\label{chiral_derivative}
\eeq
where $M^{+}_{i}$ and $M^{-}_{i}$ are monomials of chiral fields. To determine the moduli space, we first impose the vanishing of the $F$-terms, which become:
\beq
M_{i}^{+}(X_{j}) = M_{i}^{-}(X_{j}) \qquad \forall \, i ~ , 
\label{vanishing_F_terms}
\eeq
i.e. for toric phases, the vanishing $F$-term conditions are always of the form ``monomial=monomial". This property makes it possible to solve them in terms of combinatorial objects called {\it perfect matchings}. A perfect matching $p$ is a collection of edges in a brane tiling such that every node is connected to exactly one edge in $p$.

Perfect matchings can be summarized in terms of the so-called $P$-matrix, whose rows and columns are indexed by chiral fields $X_{i}$ and perfect matchings $p_\mu$, respectively. It is defined as:
\beq    
P_{i\mu} = \left\{\begin{matrix}
                            1 & \mbox{ if } & X_{i} \in p_{\mu} \\
                            0 & \mbox{ if }  & X_{i} \notin p_{\mu}
                       \end{matrix}
                        \right.     \label{def_p_matrix}
\eeq
We can think about perfect matchings as useful variables in terms of which the chiral fields in the quiver can be expressed. In particular, the following map between perfect matching variables and chiral fields:
\begin{align}
    X_{i} = \prod_{\mu}p_{\mu}^{P_{i\mu}} \label{F_term_solns}
\end{align}
automatically solves the vanishing $F$-term equations \eref{vanishing_F_terms}. Hence, there is a one-to-one correspondence between perfect matchings and GLSM fields in the toric description of the CY$_3$ moduli space. 

Further imposing the $D$-term constraints, perfect matchings are mapped to points in the corresponding $2d$ toric diagram. This description of the geometry can in general be redundant, namely multiple perfect matchings can correspond to the same point in the toric diagram. Brane tilings facilitate this process, too. Picking fundamental cycles $\gamma_{x}$ and $\gamma_{y}$ of $\mathbb{T}^2$, equivalently the boundaries of the unit cell of the brane tiling, the $\mathbb{Z}^2$ coordinates of the point in the toric diagram associated to a perfect matching $p_\mu$ are then given by:
\beq
p_{\mu} \to \sum_i P_{i \mu}  \left( \ev{X_{i},\gamma_{x}}, \ev{X_{i},\gamma_{y}} \right) ~, 
\eeq
where $\ev{X_{i},\gamma_{\alpha}}$ is the intersection paring between the edge $X_i$ and the cycle $\gamma_\alpha$. Different choices of $\gamma_{x}$ and $\gamma_{y}$ result in the same toric diagram up to $SL(2,\mathbb{Z})$ transformations.

\subsection{Brane brick models} 

\label{section_BBMs}

Similarly, the $2d$ $\mathcal{N} = (0,2)$ gauge theories on the worldvolume of D1-branes probing toric CY$_4$ singularities are fully captured by tessellations of $\mathbb{T}^{3}$ called brane brick models \cite{Franco:2015tya,Franco:2016nwv}. A brane brick model is a brane configuration involving an NS5-brane wrapping a holomorphic surface $\Sigma$ from which D4-branes are suspended, and it is related to the D1-branes at a CY$_4$ via T-duality.\footnote{As for brane tilings, we will interchangeably refer to the full brane configuration and its ``skeleton" as a brane brick model.} $\Sigma$ is the zero locus of the Newton polynomial of the CY$_4$ toric diagram. 

The gauge theory associated to a brane brick model is determined as follows. Bricks, i.e. the 3-polytopes in the tessellation, correspond to unitary gauge group. $2d$ faces represent matter fields in the bifundamental or adjoint representation of the bricks they separate.  Oriented and unoriented faces correspond to chiral and Fermi fields, respectively. Finally, every edge represents a term in the superpotential (the $m=2$ case in \eref{W_different_m}), which is given by the gauge invariant product of the chiral fields and the single Fermi field (or its conjugate) that meet at the edge.\footnote{It is possible for more than one Fermi to coincide at an edge. We refer to \cite{Franco:2016nwv} for a discussion of such cases.} 

\begin{figure}[t]
	\centering
	\includegraphics[width=12.5cm]{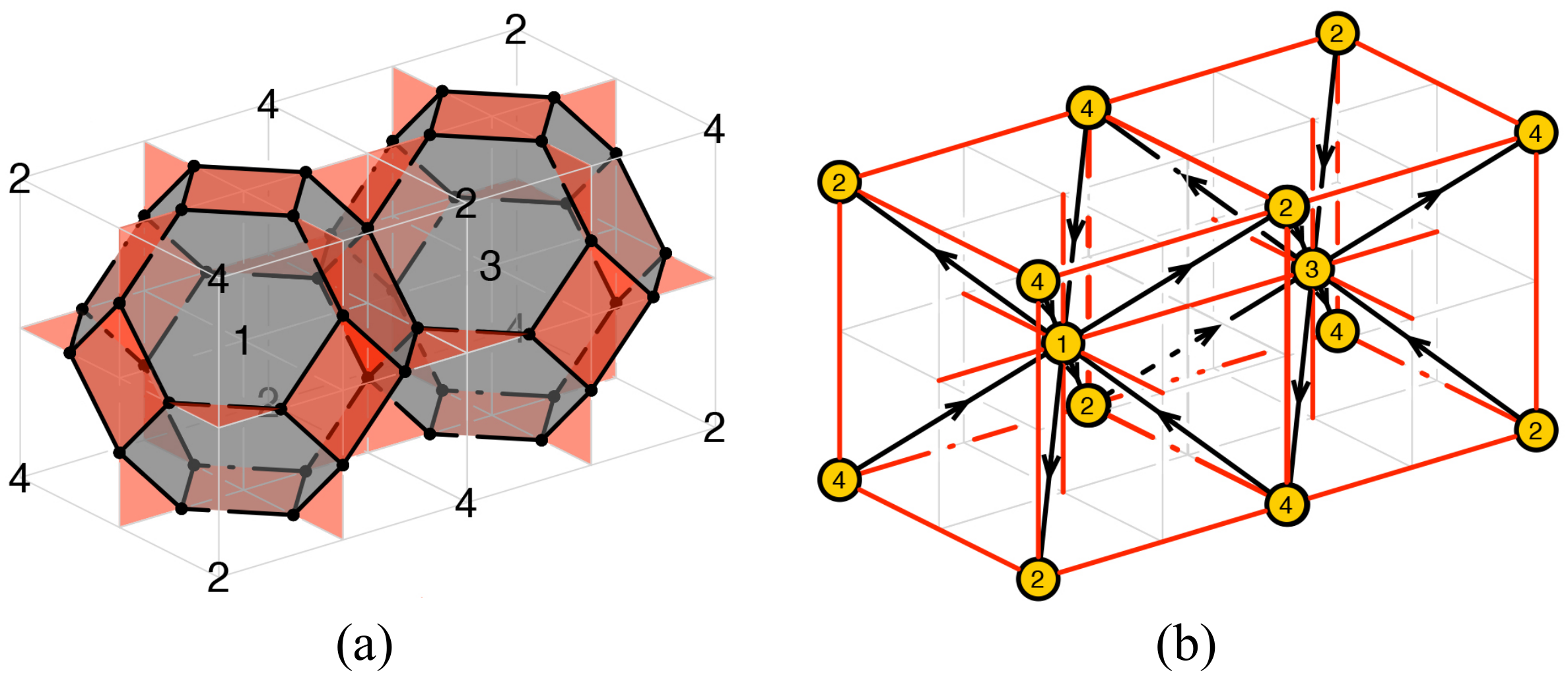}
\caption{a) Brane brick mode and b) dual periodic quiver for local $\mathbb{CP}^3$. }
	\label{BBM_periodic_quiver_C4Z4}
\end{figure}

$2d$ $\mathcal{N}=(0,2)$ theories are symmetric under the exchange of any Fermi with its conjugate. This is the symmetry between degree $m/2$ fields and their conjugates for even $m$ (in this case $m=2$) discussed in \sref{section_graded_quivers}. This symmetry is accompanied by the exchange of the corresponding $J$- and $E$-terms. The distinction between a Fermi and its conjugate, and as a result the distinction between $J$- and $E$- terms, is therefore a matter of convention. This symmetry is reflected by the fact that Fermi faces in brane brick models are unoriented. Brane brick models are dual to periodic quivers on $\mathbb{T}^3$, which contain the same information. Further details can be found in \cite{Franco:2015tya,Franco:2016nwv}. \fref{BBM_periodic_quiver_C4Z4} shows an example of brane brick model and periodic quiver, which correspond to local $\mathbb{CP}^3$ \cite{Franco:2016nwv}. Grey and red faces correspond to chiral and Fermi fields, respectively. 

The superpotential of the $2d$ $\mathcal{N}=(0,2)$ gauge theories associated to toric CY$_4$'s satisfy a toric condition \cite{Franco:2015tna}, which in this case means that every Fermi field belongs to exactly two $J$-terms and two $E$-terms with opposite signs.\footnote{Here we refer to the full gauge invariant terms, i.e. including the Fermis or their conjugates, as $J$- and $E$-terms.} The superpotential hence takes the general form
\begin{align}
    W = \sum_{a}\Lambda_{a}(J_{a}^{+}(X_{i}) - J_{a}^{-}(X_{i})) + \bar{\Lambda}_{a}(E_{a}^{+}(X_{i}) - E_{a}^{-}(X_{i}) ) ~ , 
\label{2d_toric_superpotential}
\end{align}  
where the sum runs over all Fermi fields $\Lambda_{a}$, and $J^{\pm}_{a}$ and $E^{\pm}_{a}$ are holomorphic monomials in chiral fields. In brane brick models, all Fermi faces are square. This implements the toric condition, since it implies that every Fermi fields participates in four superpotential terms, in agreement with \eref{2d_toric_superpotential}. Two of these terms correspond to $J$-terms while the two others correspond to $E$-terms. 

The toric CY$_4$ arises as the classical moduli space of the gauge theory, which can be determined in two stages. First, we impose vanishing of the chiral part of the $J$- and $E$-terms. Due to the toric form of the superpotential \eref{2d_toric_superpotential}, these conditions are once again of the form ``monomial=monomial":
\beq
    J_{a}^{+}(X_{i}) = J_{a}^{-}(X_{i}) \qquad \qquad E_{a}^{+}(X_{i}) = E_{a}^{-}(X_{i}) ~,
\label{vanishing_J_and_E}
\eeq
which allows us to solve them combinatorially. To do so, we introduce {\it brick matchings}, which are the brane brick model analogues of perfect matchings \cite{Franco:2015tya}. A brick matching $p$ is a collection of chiral and Fermi fields such that:
    \begin{itemize}
        \item 
            For every Fermi field $\Lambda_{a}$, $p$ contains exactly either $\Lambda_{a}$ or $\bar{\Lambda}_{a}$.
        \item
            If $p$ contains $\Lambda_{a}$, it contains exactly one chiral field in each of $E_a^+$ and $E_a^-$.
        \item
            If $p$ contains $\bar{\Lambda}_{a}$, it contains exactly one chiral field in each of $J_a^+$ and $J_a^-$.
    \end{itemize}
    We can summarize the chiral field content of brick matchings by means of the $P$-matrix, defined as in \eqref{def_p_matrix}.\footnote{The $P$-matrix can be extended to include the Fermi field content of brick matchings \cite{Franco:2015tya}. This extra information has various applications but it is not necessary for the determination of the moduli space.} Again, the map between chiral fields and brick matchings given by equation \eqref{F_term_solns} solves \eref{vanishing_J_and_E}.

The final step consists of imposing the vanishing of $D$-terms. As for brane tilings, there is an alternative way of finding the $\mathbb{Z}^{3}$ coordinates in the toric diagram for every brick matching. They are given by: 
\beq
p_{\mu} \to  \sum_i P_{i \mu}  \left( \ev{X_{i},\gamma_{x}}, \ev{X_{i},\gamma_{y}} , \ev{X_{i},\gamma_{z}}\right) ~,
\eeq    
where $\gamma_{x}$, $\gamma_{y}$ and $\gamma_{z}$ are the edges of the unit cell of the brane brick model. Again, different choices of $\gamma_\alpha$ lead to the same toric diagram up to $SL(3,\mathbb{Z})$ transformations.

\section{Generalized dimer models}
 
 \label{section_generalized_dimers}
 
We have reviewed brane tilings and brane brick models and discussed how they simplify the relation between geometry and gauge theory. In this section we will describe an infinite generalization of them, which streamline the connection between toric CY $(m+2)$-folds and $m$-graded quivers.

\subsection{Toric quivers}
 
 \label{section_toric_quivers}     
 
The CY$_{m+2}$ associated to an $m$-graded quiver arises as its {\it classical moduli space}. Extending the usual notion for $m\leq 3$, it is defined as the center of the Jacobian algebra with respect to fields of degree $m-1$ \cite{Franco:2017lpa}. This corresponds to imposing the relations:
\beq
{\partial W \over \partial \Phi^{(m-1)}}=0~, \ \ \ \ \ \forall \, \Phi^{(m-1)}
\label{relations}
\eeq
plus gauge invariance.\footnote{For $m=2$, the fields $\Phi^{(1)}$ in \eref{relations} denote both $\Phi^{(1)}$ and $\bar\Phi^{(1)}$, namely the Fermi and conjugate Fermi fields in the $2d$ $\mathcal{N}=(0,2)$ gauge theory.}

Since the superpotential has degree $m-1$, the terms that contribute to the relations in \eref{relations} are of the general form $\Phi^{(m-1)} P(\Phi^{(0)})$, where $P(\Phi^{(0)})$ is a holomorphic function of chiral fields. We will refer to such terms as $J$-terms. The resulting relations \eref{relations} then consist entirely of chiral fields. 

The quivers theories associated to toric singularities are endowed with additional structure. Their global symmetry contains a $U(1)^{m+2}$ Cartan subgroup, coming from the isometries of the underlying CY$_{m+2}$.

Furthermore, for every toric CY$_{m+2}$ there exists at least one {\it toric phase}, which is a theory satisfying the following properties.\footnote{Acting on this theory with sequences of mutations, we can generate other phases, both toric (if they exist) and non-toric.} First, for $N$ regular branes and no fractional branes, the ranks of all gauge groups are equal to $N$. In other words, there exists a solution to the anomaly cancellation condition \eref{anomaly_odd} or \eref{anomaly_even} in which the ranks of all gauge groups are equal and unconstrained. 

In addition, the superpotential of a toric phase obeys a {\it toric condition}, according to which every field of degree $m-1$ appears in exactly two $J$-terms, with opposite signs. Namely,
\beq
W= \Phi^{(m-1)}_a J_a^+(\Phi^{(0)})- \Phi_a^{(m-1)} J_a^-(\Phi^{(0)}) + \ldots~,
\label{general_toric_W}
\eeq
where dots indicate terms that do not contain $\Phi_a^{(m-1)}$. This condition generalizes the toric conditions for the $m=1,2$ cases discussed in \sref{section_brane_tilings} and \sref{section_BBMs}. The relations \eref{relations} then take the ``monomial=monomial" form:
\beq
J_a^+(\Phi^{(0)}) = J_a^-(\Phi^{(0)}) ~.
\label{general_toric_J}
\eeq
This property is of central importance for the generalized dimers and the associated combinatorial tools that we will introduce later. A straightforward way of deriving the toric condition is as follows. It is satisfied by the $\mathbb{C}^{m+2}$ quivers as it will be explained in \sref{section_orbifolds_Cm+2}, it is inherited by its $\mathbb{C}^{m+2}/(\mathbb{Z}_{N_1} \times \cdots \times \mathbb{Z}_{N_{m+1}})$ orbifolds and it is preserved by partial resolution, with which we can reach an arbitrary toric CY$_{m+2}$.

As we previously mentioned, in this paper we will exclusively focus on toric phases, so we will no longer emphasize this distinction.

\subsection{Generalized dimer models and periodic quivers}
 
We are now ready to introduce {\it generalized dimers of order $m$}, or {\it $m$-dimers} for short, which fully encode the $m$-graded quivers with superpotentials of toric phases associated to toric CY$_{m+2}$'s and simplify their connection to geometry.

Consider the Newton polynomial of the toric CY$_{m+2}$ under consideration, which is given by: 
\beq
P(x_1,\ldots, x_{m+1})=\sum_{\vec{v} \in V} c_{\vec{v}} \, x_1^{v_1} \ldots x_{n-1}^{v_{m+1}} ~ , 
\label{Newton_polynomial}
\eeq
where $x_\mu\in \mathbb{C^*}$, $\mu=1,\ldots,m+1$, the $c_{\vec{v}}$ are complex coefficients and the sum runs over points $\vec{v}$ in the toric diagram. By rescaling the $x_\mu$'s, it is possible to set $m+2$ of the coefficients to 1. The freedom in the remaining coefficients captures dual phases of the quiver theory.

In addition, let us consider the {\it coamoeba projection} from $(\mathbb{C^*})^{m+1}$ to $\mathbb{T}^{m+1}$: 
\beq
(x_1,\ldots, x_{m+1}) \mapsto (\arg(x_1),\ldots, \arg(x_{m+1})) ~ .
\eeq

We define an $m$-dimer as the coamoeba projection of the holomorphic surface $\Sigma_m$, which in turn is given by the zero locus of the Newton polynomial:
\beq
\Sigma_m: \ \ \ \ P(x_1,\ldots, x_{m+1}) = 0 ~ .
\eeq 
This definition reproduces the $m\leq 3$ cases (elliptic models, brane tilings, brane brick models and brane hyperbrick models) and naturally generalizes them. As usual, most of the time we will focus on its tropical limit or ``skeleton", which is a tessellation of $\mathbb{T}^{m+1}$.\footnote{For this reason, these objects have been dubbed {\it tropical coamoebas} in the mathematical literature \cite{2007arXiv0710.1898I}.}    

The quiver theory can be read from the $m$-dimer as follows. Every codimension-$0$ face $i$, which we will denote brick, corresponds to a gauge group. Every codimension-1 face common to bricks $i$ and $j$ has an orientation and a degree $c$, $0 \le c \le m$, and corresponds to a bifundamental (or an adjoint field if $i=j$) field of degree $c$. As usual, we can flip the orientation by conjugation, which simultaneously changes the degree to $m-c$. Codimension-$2$ faces are such that the degree of faces they bound sum to $m-1$ and map to superpotential terms. 

While determining the tessellation is relatively straightforward, establishing the orientations and degrees of its codimension-1 faces is not.  For $m=2$, this problem has been addressed in \cite{Franco:2016qxh}, but an algorithm for general $m$ is still lacking.  In practice, there are efficient alternatives for approaching this problem. One of them is obtaining the theory for the desired geometry by partial resolution of an orbifold. Implementing such partial resolution is considerably simplified using $m$-dimers.

Clearly, the structure of $m$-dimers is richer than what we have exploited so far. In particular, starting at $m=2$, it is natural to ask whether faces of codimension higher than 2 have a gauge theory interpretation. It is tempting to speculate that they are connected to the $A_\infty$ relations among multi-products in the quiver algebra \cite{Closset:2017yte,Closset:2018axq}. We plan to revisit this question in future work.   

Via graph dualization, $m$-dimers are in one-to-one correspondence with {\it periodic $m$-quivers} in $\mathbb{T}^{m+1}$. Both objects contain exactly the same information. Periodic quivers not only summarize the gauge symmetry and field content. They are such that every minimal plaquette, namely the duals to codimension-2 faces of the $m$-dimer, corresponds to a term in the superpotential. Table \ref{table_dictionary} summarizes the map between $m$-dimers and periodic $m$-quivers.

\begin{table}[h]
\centering
\begin{tabular}{|c|c|}
\hline
{\bf m-dimer} & {\bf Periodic m-quiver} 
\\
\hline\hline
Codimension-0 face (brick) & Gauge group \\ \hline
 \ \ \ Codimension-1 face of degree $c$ \ \ \ & \ \ \ Degree $c$ field in the bifundamental \ \ \ \\
between bricks $i$ and $j$ &  representation of nodes $i$ and $j$ \\
& (adjoint for $i$ = $j$) \\ \hline
Codimension-2 face & Plaquette encoding a monomial in \\
& the superpotential  \\
\hline
\end{tabular}
\caption{Dictionary mapping $m$-dimers to periodic $m$-quivers (equivalently, toric $m$-quivers with superpotential).
\label{table_dictionary}
}
\end{table}

It is convenient to decompose the $U(1)^{m+2}$ Cartan subgroup of the global symmetry as $U(1)^{m+1}_{flavor} \times U(1)_R$. The $U(1)^{m+1}_{flavor}$ is nicely geometrized by $m$-dimers and periodic quivers, where it is mapped to the fundamental directions of $\mathbb{T}^{m+1}$. 
              
For $m\geq 3$, $m$-dimers and periodic quivers have more than three dimensions and hence become difficult to visualize. Their structure can be captured by various projections, such as the tomography of \cite{Futaki:2014mpa,Franco:2016qxh}. However, as we will show in this paper, several powerful tools follow from the existence of $m$-dimers, equivalently from the structure of toric phases, and do not require their explicit visualization. For simplicity, we will often phrase our discussions in terms of periodic quivers.

\section{$\mathbb{C}^{m+2}$ and permutohedra}

\label{section_Cm+2_permutohedra}

We now discuss the $m$-dimers associated to flat space, $\mathbb{C}^{m+2}$. These dimers can be regarded as the simplest ones but also as the most universal, since the ones for any other toric CY$_{m+2}$ can be obtained from their orbifolds, which in turn correspond to stacking multiple copies of the same brick, by partial resolution. 

\begin{figure}[t]
	\centering
	\includegraphics[width=12cm]{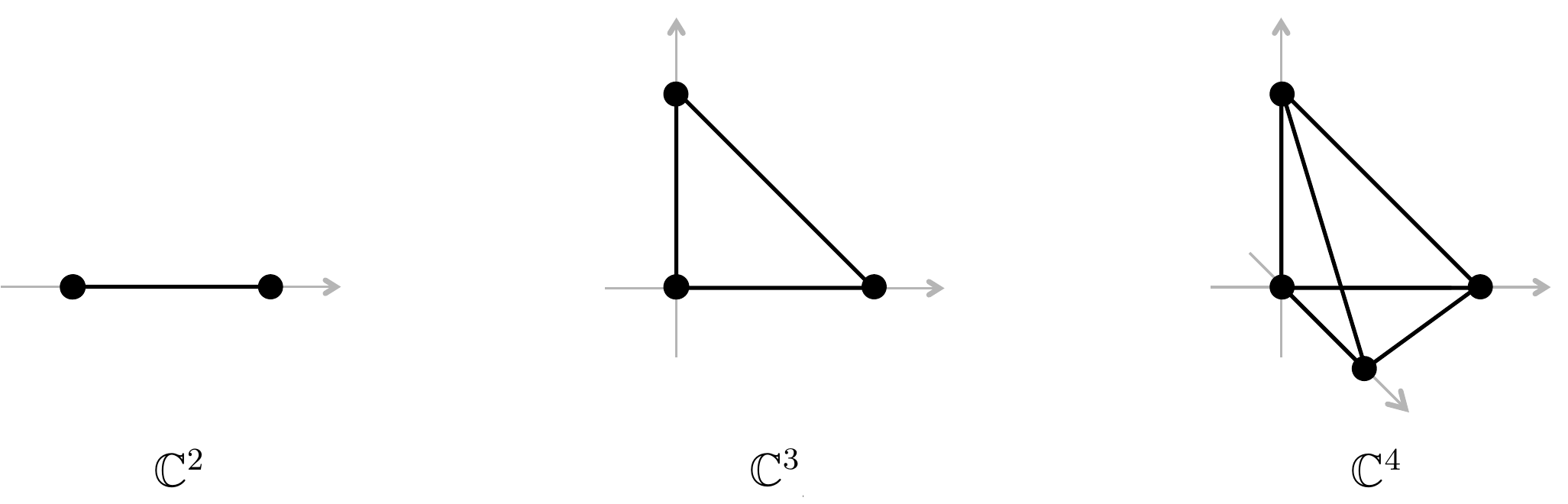}
\caption{Toric diagrams for $\mathbb{C}^{m+2}$ with $m=0,1,2$.}
	\label{toric_diagrams_Cn}
\end{figure}

The toric diagram for $\mathbb{C}^{m+2}$ is the minimal simplex in $\mathbb{Z}^{m+1}$, namely it is given by the following points:
\beq
\begin{array}{l}
v_0=(0,\ldots,0)~, \cr
v_1= (1,0,0,\ldots,0)~, \quad
 v_2=(0,1,0,\ldots,0)~, \quad \ldots~, \quad
 v_{m+1}=(0,0,\ldots,0, 1)~.
\end{array}
\eeq
\fref{toric_diagrams_Cn} shows the toric diagrams for $m\leq 3$. The geometry has an $SU(m+2)$ isometry, which maps to an $SU(m+2)$ global symmetry of the corresponding quiver theories.

\subsection{Quiver theories}

This infinite family of theories was first discussed in full generality in \cite{Closset:2018axq} where it was independently derived using both the algebraic dimensional reduction procedure introduced in \cite{Closset:2018axq} and the topological B-model. We quickly review it here. For $m=0, 1, 2, 3$ these theories correspond to maximally supersymmetric Yang-Mills in $d=6,4,2,0$. For general $m$, the quiver is defined as follows:
\begin{itemize}
          \item It has a single node.
          \item It contains adjoint fields $\Phi^{(c,c+1)}$ of degree $0\leq c \leq \floor{{m\over 2}}$. Here we have introduced a notation with two superindices, in which $\Phi^{(c;k)}$ indicates an arrow of degree $c$ in the $k$-index totally antisymmetric representation of the global $SU(m+2)$ symmetry. Then, every field $\Phi^{(c,c+1)}$ transforms in the antisymmetric $(c+1)$-index representation of $SU(m+2)$. The conjugates of these fields $\overline{\left( \Phi^{(c,c+1)} \right)} \equiv \bar{\Phi}^{(m-c;m+1-c)}$ have degree $m-c$ and transform in the antisymmetric $(m+1-c)$-index representation of $SU(m+2)$.
          \item In the case of even $m$, the multiplicity of the unoriented degree-${m\over 2}$ fields is half the dimension of the corresponding representation. The full representation can be built in terms of $\Phi^{({m\over 2})}$ and $\bar{\Phi}^{({m\over 2})}$.
      \end{itemize}

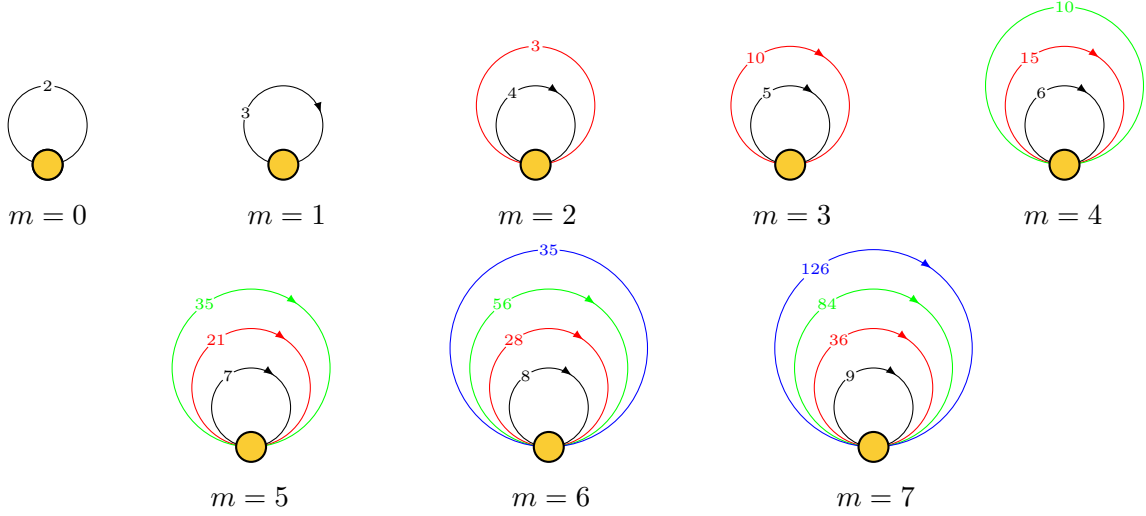
\begin{figure}
\captionsetup[subfigure]{labelformat=empty}
        \centering
        \begin{subfigure}[b]{0.11\textwidth}
             \newcommand{\pos}{0.5}
             \newcommand{\arrowHeadPosition}{0.7} 
            \begin{tikzpicture}[scale=1.75 , decoration={markings,mark=at position \arrowHeadPosition with {\arrow{latex}}}]
                \tikzstyle{every node}=[circle,thick,fill=yellow2,draw,inner sep=4pt,font=\tiny]
                \draw (0,0) node (A) {};
                    \draw[black] (0,0) arc(270:-90:0.3) node[pos = \pos, draw = none , fill = white , inner sep = 0]{$2$};
                \draw (0,0) node (A) {};
            \end{tikzpicture} 
            \subcaption{$m = 0$ \ \ \ \ \ }        
        \end{subfigure}
        \hspace{0.07\textwidth}
        \begin{subfigure}[b]{0.115\textwidth}
            \newcommand{\pos}{0.3}
            \newcommand{\arrowHeadPosition}{0.7} 
            \begin{tikzpicture}[scale=1.75 , decoration={markings,mark=at position \arrowHeadPosition with {\arrow{latex}}}]
                \tikzstyle{every node}=[circle,thick,fill=yellow2,draw,inner sep=4pt,font=\tiny]
                    \draw[postaction={decorate}, black] (0,0) arc(270:-90:0.3) node[pos = \pos, draw = none , fill = white , inner sep = 0]{$3$};
                \draw (0,0) node (A) {};
            \end{tikzpicture} 
             \subcaption{$m = 1$ \ \ \ \ }
        \end{subfigure}
        \hspace{0.07\textwidth}
        \begin{subfigure}[b]{0.13\textwidth}
            \newcommand{\pos}{0.4}
            \newcommand{\arrowHeadPosition}{0.6} 
            \begin{tikzpicture}[scale=1.75 , decoration={markings,mark=at position \arrowHeadPosition with {\arrow{latex}}}] 
                \tikzstyle{every node}=[circle,thick,fill=yellow2,draw,inner sep=4pt,font=\tiny]
                \draw[postaction={decorate}, black] (0,0) arc(270:-90:0.3) node[pos = \pos, draw = none , fill = white , inner sep = 0]{$4$};
                \draw[red] (0,0) arc(270:-90:0.44999999999999996) node[pos = 0.5, draw = none , fill = white , inner sep = 0]{$3$};
                \draw (0,0) node (A) {};
            \end{tikzpicture} 
             \subcaption{$m = 2$ \ \ \ }
        \end{subfigure}
        \hspace{0.07\textwidth}
        \begin{subfigure}[b]{0.13\textwidth}
            \newcommand{\pos}{0.4}
            \newcommand{\arrowHeadPosition}{0.6} 
            \begin{tikzpicture}[scale=1.75 , decoration={markings,mark=at position \arrowHeadPosition with {\arrow{latex}}}] 
                \tikzstyle{every node}=[circle,thick,fill=yellow2,draw,inner sep=4pt,font=\tiny]
                \draw[postaction={decorate}, black] (0,0) arc(270:-90:0.3) node[pos = \pos, draw = none , fill = white , inner sep = 0]{$5$};
                \draw[postaction={decorate}, red] (0,0) arc(270:-90:0.44999999999999996) node[pos = \pos, draw = none , fill = white , inner sep = 0]{$10$};
                \draw (0,0) node (A) {};
            \end{tikzpicture} 
             \subcaption{$m = 3$ \ \ \ }
        \end{subfigure}
        \hspace{0.07\textwidth}
        \begin{subfigure}[b]{0.15\textwidth}
            \newcommand{\pos}{0.4}
            \newcommand{\arrowHeadPosition}{0.6} 
            \begin{tikzpicture}[scale=1.75 , decoration={markings,mark=at position \arrowHeadPosition with {\arrow{latex}}}]
                \tikzstyle{every node}=[circle,thick,fill=yellow2,draw,inner sep=4pt,font=\tiny]
                \draw[postaction={decorate}, black] (0,0) arc(270:-90:0.3) node[pos = \pos, draw = none , fill = white , inner sep = 0]{$6$};
                \draw[postaction={decorate}, red] (0,0) arc(270:-90:0.44999999999999996) node[pos = \pos, draw = none , fill = white , inner sep = 0]{$15$};
                \draw[green] (0,0) arc(270:-90:0.6) node[pos = 0.5, draw = none , fill = white , inner sep = 0]{$10$};
                \draw (0,0) node (A) {};
            \end{tikzpicture} 
             \subcaption{$m = 4$ \ \ }
        \end{subfigure}
        \begin{subfigure}[b]{0.15\textwidth}
            \newcommand{\pos}{0.4}
            \newcommand{\arrowHeadPosition}{0.6} 
            \begin{tikzpicture}[scale=1.75 , decoration={markings,mark=at position \arrowHeadPosition with {\arrow{latex}}}] 
                \tikzstyle{every node}=[circle,thick,fill=yellow2,draw,inner sep=4pt,font=\tiny]
                \draw[postaction={decorate}, black] (0,0) arc(270:-90:0.3) node[pos = \pos, draw = none , fill = white , inner sep = 0]{$7$};
                \draw[postaction={decorate}, red] (0,0) arc(270:-90:0.44999999999999996) node[pos = \pos, draw = none , fill = white , inner sep = 0]{$21$};
                \draw[postaction={decorate}, green] (0,0) arc(270:-90:0.6) node[pos = \pos, draw = none , fill = white , inner sep = 0]{$35$};
                \draw (0,0) node (A) {};
            \end{tikzpicture} 
             \subcaption{$m = 5$ \ \ }
        \end{subfigure}
        \hspace{0.07\textwidth}
        \begin{subfigure}[b]{0.19\textwidth}
            \newcommand{\pos}{0.4}
            \newcommand{\arrowHeadPosition}{0.6} 
            \begin{tikzpicture}[scale=1.75 , decoration={markings,mark=at position \arrowHeadPosition with {\arrow{latex}}}] 
                \tikzstyle{every node}=[circle,thick,fill=yellow2,draw,inner sep=4pt,font=\tiny]
                \draw[postaction={decorate}, black] (0,0) arc(270:-90:0.3) node[pos = \pos, draw = none , fill = white , inner sep = 0]{$8$};
                \draw[postaction={decorate}, red] (0,0) arc(270:-90:0.44999999999999996) node[pos = \pos, draw = none , fill = white , inner sep = 0]{$28$};
                \draw[postaction={decorate}, green] (0,0) arc(270:-90:0.6) node[pos = \pos, draw = none , fill = white , inner sep = 0]{$56$};
                \draw[blue] (0,0) arc(270:-90:0.75) node[pos = 0.5, draw = none , fill = white , inner sep = 0]{$35$};
                \draw (0,0) node (A) {};
            \end{tikzpicture} 
             \subcaption{$m = 6$ \ \ }
        \end{subfigure}
        \hspace{0.07\textwidth}
        \begin{subfigure}[b]{0.19\textwidth}
            \newcommand{\pos}{0.4}
            \newcommand{\arrowHeadPosition}{0.6} 
            \begin{tikzpicture}[scale=1.75 , decoration={markings,mark=at position \arrowHeadPosition with {\arrow{latex}}}] 
                \tikzstyle{every node}=[circle,thick,fill=yellow2,draw,inner sep=4pt,font=\tiny]
                \draw[postaction={decorate}, black] (0,0) arc(270:-90:0.3) node[pos = \pos, draw = none , fill = white , inner sep = 0]{$9$};
                \draw[postaction={decorate}, red] (0,0) arc(270:-90:0.44999999999999996) node[pos = \pos, draw = none , fill = white , inner sep = 0]{$36$};
                \draw[postaction={decorate}, green] (0,0) arc(270:-90:0.6) node[pos = \pos, draw = none , fill = white , inner sep = 0]{$84$};
                \draw[postaction={decorate}, blue] (0,0) arc(270:-90:0.75) node[pos = \pos, draw = none , fill = white , inner sep = 0]{$126$};
                \draw (0,0) node (A) {};
            \end{tikzpicture} 
             \subcaption{$m = 7$ \ \ }
        \end{subfigure}
        \caption{Quivers for $\mathbb{C}^{m+2}$. The multiplicities of fields, i.e. the dimensions of the representations for the $SU(m+2)$ global symmetry, are indicated on the arrows. For $m$ even, the multiplicity of the outmost (unoriented) line is half the dimension of the corresponding representation. 
        Black, red, green, blue and purple arrows represent fields of degree 0, 1, 2, 3 and 4, respectively.
        \label{quivers_Cn}}
    \end{figure}

\paragraph{Superpotential.}
The superpotential can be written compactly by exploiting the $SU(m+2)$ global symmetry. It is:
\beq
W = \sum_{i+j+k = m+2}\Phi^{(j-1;j)}\Phi^{(k-1;k)}\bar{\Phi}^{(m+1-j-k;m+2-j-k)} ~ .
            \label{potential_Cn}
\eeq
Every term has $m+2$ $SU(m+2)$ flavor indices, which are contracted with a Levi-Civita tensor to form an $SU(m+2)$ invariant. We have suppressed them in the interest of a cleaner notation.

\paragraph{Periodic quiver.} 
We now discuss how periodic quivers neatly capture the $\mathbb{C}^{m+2}$ quivers and their superpotentials. While visualizing periodic quivers beyond $m=2$ is challenging, they can be described fairly straightforwardly.

The periodic quiver can be embedded in the torus $\mathbb{T}^{m+1}\equiv \mathbb{R}^{m+1} \, \rm{mod} \, (\mathbb{Z}^{m+1})$. The unit cell thus becomes the domain $[0,1]^{m+1}$. We locate the single node of the quiver at the origin.  

Let us now consider the arrows. There is a field stretching from the origin to every corner of the unit cell. Since all the corners are identified, these are adjoint fields. There are $2^{m+1}-1$ corners other than the origin, which is indeed the total number of fields in the quiver.

The degree $c$ of the field connecting the origin to the corner with coordinates $q_{\alpha}$ is given by the $L^{1}$ norm of the point minus one, i.e.:
      \begin{align}
           c = \sum_{\alpha=1}^{m+1}\abs{q_{\alpha}} -1 ~.
      \end{align} 
Hence, considering only fields outgoing from the origin, there are $\binom{m+1}{c+1}$ fields of degree $c$ and $\binom{m+1}{m+1-c}$ of degree $m-c$, whose conjugates also have degree $c$. If $c \ne m-c$ the two sets are distinct and the total number of fields of degree $c$ is
      \begin{align}
           \binom{m+1}{c+1} + \binom{m+1}{c} = \binom{m+2}{c+1} ~ .
      \end{align} 
This is precisely the dimension of the antisymmetric $(c+1)$-index representation of $SU(m+2)$. When $c =m/2$ the two sets coincide and the number of fields of degree $c$ is half the dimension of the $(m/2+1)$-index representation of $SU(m+2)$. In summary, this construction of the periodic quivers nicely reproduces the quivers for $\mathbb{C}^{m+2}$ introduced above. 

Finally, it can be verified that going around any collection of three corners of the unit cell gives rise to a minimal plaquette of degree $m-1$, as required. This reproduces the cubic superpotential in \eref{potential_Cn}.

\subsection{m-dimers}

The $m$-dimer for $\mathbb{C}^{m+2}$ takes a remarkably elegant form. It consist of a tiling of $\mathbb{T}^{m+1}$ by a single brick, which is a {\it permutohedron} of order $(m+2)$, or $(m+2)$-permutohedron for short. This fact was originally noted by Futaki and Ueda in their seminal paper \cite{Futaki:2014mpa}.\footnote{We thank Eduardo Garcia-Valdacasas, who independently arrived to this conclusion from an analysis of the corresponding quiver theory, for sharing this insight with us.} The $(m+2)$-permutohedron is an $(m+1)$-dimensional polytope embedded in $(m+2)$-dimensions. The coordinates of its vertices are the permutations of the set $\{1,\ldots, m+2 \}$. The number of vertices is thus $(m+2)!$, each of which is adjacent to $m+1$ others. Every edge connects two vertices that are related by exchanging two coordinates, the values of which differ by one.

More generally, the $(m+2)$-permutohedron has a facet for every non-empty proper subset of $\{1,\ldots, m+2 \}$. The number of codimension-$d$ facets is:
\beq
F(m,d)=(d+1)! \, S(m+2,d+1) ~,
\label{facets_permutohedron}
\eeq
where $S(i,j)$ denotes the Stirling numbers of the second kind. 

For example, the $m$-permutohedra for $m=1,2,3,4$ are the line segment, hexagon, truncated octahedron and omnitruncated 5-cell, respectively. The first three objects in this list are well known from the study of elliptic models, brane tilings and brane brick models. 

Let us focus on codimension-1 and 2 facets, which correspond to fields and superpotential terms. Using \eref{facets_permutohedron}, we get the table on the left of \eref{fields_Wterms_permutohedron}. Since for $\mathbb{C}^{m+2}$ we have a single brick with periodic identifications, the number of codimension-$d$ facets must be divided by $d$, giving rise the table on the right. These results are in perfect agreement with the corresponding quiver theories. Indeed, these $m$-dimers are connected to the periodic quivers discussed in the previous section by graph dualization.
\beq
\begin{array}{|c|c|c|}
\cline{2-3} 
\multicolumn{1}{c|}{} &   \multicolumn{2}{c|}{\mbox{Codimension}} \\ \hline
\ m \ \ & \ \ \ \ 1 \ \ \ \ \ & \ \ \ \ 2 \ \ \ \ \ \\ \hline \hline
0 & 2 & - \\
1 & 6 &  6 \\
2 & 14 & 36  \\
3 & 30 & 150  \\
4 & 62 & 540 \\
5 & 126 & 1806 \\
6 & 254 & 5796  \\ \hline
\end{array} 
\qquad \to \qquad
\begin{array}{|c|c|c|}
\cline{2-3} 
\multicolumn{1}{c|}{} &   \multicolumn{2}{c|}{\mbox{Codimension}} \\ \hline
\ m \ \ & \ \ \ \ 1 \ \ \ \ \ & \ \ \ \ 2 \ \ \ \ \ \\ \hline \hline
0 & 1 & - \\
1 & 3 &  2 \\
2 & 7 & 12  \\
3 & 15 & 50  \\
4 & 31 & 180 \\
5 & 63 & 602 \\
6 & 127 & 1932 \\ \hline
\end{array}
\label{fields_Wterms_permutohedron}
\eeq

\smallskip

\paragraph{Orbifolds and higgsing.} The $m$-dimer for a $\mathbb{C}^{m+2}/(\mathbb{Z}_{N_1} \times \cdots \times \mathbb{Z}_{N_{m+1}})$ orbifold is simply given by a $N_1\times \cdots \times N_{m+1}$ stack of $(m+2)$-permutohedra bricks. The action of the generators of the orbifold group determines the periodicity conditions in $\mathbb{T}^{m+1}$, as we will elaborate in \sref{section_orbifolds_Cm+2}.

The $m$-dimer for an arbitrary toric CY$_{m+2}$ can be obtained by starting from an orbifold whose toric diagram contains the desired one and performing partial resolution. This process translates into ``higgsing" of the quiver which, in terms of the dimer, corresponds to removing the codimension-1 faces associated to the chiral fields acquiring non-zero ``VEVs" and recombining the bricks accordingly. Pairs of fields might become massive in this process and can be integrated out.

 \section{Moduli spaces and generalized perfect matchings}
 
 \label{section_perfect_matchings}
    
As discussed in section \sref{section_toric_quivers}, the CY$_{m+2}$ corresponds to the moduli space of the $m$-graded quiver. The first step in its determination is imposing the vanishing of $J$-terms. Due to the toric condition \eref{general_toric_J}, solving these equations can be accomplished combinatorially.

\subsection{Generalized perfect matchings}

\label{section_generalized_pms}
    
We define a {\it generalized perfect matching} $p$ of an $m$-dimer as a collection of fields satisfying:
    \begin{itemize}
        \item[{\bf 1)}] $p$ contains precisely one field from each term in $W$.
        \item[{\bf 2)}] For every field $\Phi$ in the quiver, either $\Phi$ or $\bar{\Phi}$ is in $p$. 
    \end{itemize}
In what follows, we will drop the unwieldy `generalized' when it can be inferred from the context and just call these objects perfect matchings.

It is important to emphasize that while perfect matchings have a natural interpretation in terms of $m$-dimers, they can be defined, as we have just done, purely in terms of the quiver theories.

\paragraph{Vanishing of $J$-terms.}
Perfect matchings as defined above allow us to solve \eref{general_toric_J} combinatorially. The process is essentially the same as in the case of brane tilings and brane brick models. We encode the relation among chiral fields and perfect matchings in terms of the $P$-matrix:\footnote{As we mentioned for the $m=2$ case in \sref{section_BBMs}, the $P$-matrix can be extended to include the fields of all degrees that form a perfect matching. We expect that this extended information will be useful for studying various structures associated to $m$-dimers which are yet to be discovered, but it is not necessary for computing the moduli space.}
\beq
            P_{i,\mu} = \left\{\begin{matrix}
                                    1 & \mbox{ if } & \Phi_{i}^{(0)} \in p_{\mu} \\
                                    0 & \mbox{ if }  & \Phi_{i}^{(0)} \notin p_{\mu}
                               \end{matrix}
                                \right.     \label{def_p_matrix2}
\eeq
Using the $P$-matrix we define the map between perfect matchings $p_{\mu}$ and chiral fields $\Phi^{(0)}_{i}$ as follows:
\beq
\Phi^{(0)}_{i} = \prod_{\mu}p_{\mu}^{P_{i,\mu}} \label{F_term_solns2} ~.
\eeq
With this map, the vanishing of $J$-terms \eref{general_toric_J} becomes
        \begin{align}
            \prod_{\Phi_{i}^{(0)} \in J_{a}^{+}}\prod_{p_\mu}p_{\mu}^{P_{i,\mu}} = \prod_{\Phi_{i}^{(0)} \in J_{a}^{-}}\prod_{p_\mu}p_{\mu}^{P_{i,\mu}} ~.
        \end{align}
Remarkably, the definition of perfect matchings introduced above is such that these equations are always satisfied. This is because for every $p_\mu$, either $p_\mu$ has no chiral field in both $J_{a}^{+}$ and $J_{a}^{-}$ or else it has exactly one chiral field in each of them. This is clearly the case since the relevant terms in the superpotential are $\Phi^{(m-1)}_a (J_a^+ - J_a^-)$ and a perfect matching picks exactly one field from each of these two terms. Hence it either contains $\Phi^{(m-1)}_a$ and no chirals from $J^{+}_{a}$ and $J^{-}$, or it does not contain $\Phi^{(m-1)}_a$ but involves one chiral from each of $J^{+}_{a}$ and $J^{-}$.

\paragraph{From perfect matchings to the toric diagram.} 
The second step in the computation of moduli space is to assign positions in the integer lattice $\mathbb{Z}^{m+1}$ to the perfect matchings, in order to construct the toric diagram of the CY$_{m+2}$. We can do so by a straightforward generalization of the procedure we previously outlined for brane tilings and brane brick models.  We pick fundamental cycles $\gamma_{\alpha}$, $\alpha=1,\ldots, m+1$, of the torus $\mathbb{T}^{m+1}$ in which the $m$-dimer is embedded. Since chiral fields are oriented codimension-1, faces we can define the intersection pairing between the chiral fields and the fundamental cycles:
\beq
        \ev{\Phi_{i}^{(0)},\gamma_{\alpha}} = \left\{\begin{array}{cl}
                                                  \pm 1 & \mbox{ if }  \Phi_{i}^{(0)} \mbox{ intersects } \gamma_{\alpha} \\
                                                  0 & \mbox{ if }   \Phi_{i}^{(0)} \mbox{ does not intersect } \gamma_{\alpha}
                                               \end{array}
                                              \right.  
\eeq
Recall that degree-$m$ fields should be regarded as conjugate chirals.

The position of a perfect matching in the toric diagram is then given by:
\beq
p_{\mu} \to  \sum_i P_{i \mu}  \left(\ev{\Phi_{i}^{(0)},\gamma_1}, \cdots , \ev{\Phi_{i}^{(0)},\gamma_{m+1}}\right) ~.
\eeq    
Alternative choices of $\gamma_\alpha$ give rise to the same toric diagram up to $SL(m+1,\mathbb{Z})$ transformations.

\subsubsection{$m=1,2$ case}
Let us verify that for $m=1,2$ generalized perfect matchings indeed reduce to ordinary perfect matchings and brick matchings, respectively.

\paragraph{Ordinary perfect matchings.} 
For $m=1$, only chiral fields appear in the superpotential due to holomorphy and an ordinary perfect matchings $p$ is defined by the first of the conditions above. The second condition can be implemented by simply adding to $p$ the conjugates of the rest of the fields, since the conjugates do not appear in any superpotential term. Therefore, ordinary perfect matchings are in one-to-one correspondence with generalized perfect matchings.

\paragraph{Brick matchings.} 
For $m=2$, a brick matching $p$ is obtained by requiring the first condition but the second condition is imposed only for Fermis. Again, since the superpotential does not contain conjugate chiral fields, we can uniquely extend a brick matching to a generalized perfect matching by adding the conjugates of all the chiral fields which are not in $p$.

\subsubsection{The $m=0$ case}

Although we motivated $m$-dimers by discussing brane tilings ($m=1$) and brane brick models ($m=2$), the natural starting point is $m=0$. This case corresponds to $6d$ $\mathcal{N}=(1,0)$ supersymmetric gauge theories on the worldvolume of D5-branes probing toric CY 2-folds. 

The only toric CY 2-folds are $\mathbb{C}^2$ and its orbifolds $\mathbb{C}^{2}/\mathbb{Z}_{n}$, whose toric diagrams are given by the integer points $0,\ldots, n$. The quiver for $\mathbbm{C}^{2}/\mathbbm{Z}_{n}$ is an $n$-node necklace quiver, namely the affine Dynkin diagram $\tilde{A}_{n}$. 

Let us now discuss the perfect matchings and how they give rise to the toric diagrams. For $m=0$, all the matter fields and their conjugates have degree $0$ so the fields or edges in the periodic quiver are unoriented. There is no superpotential, since it should have degree $m-1=-1$. This implies that the first condition in the definition of perfect matchings is trivially satisfied. A perfect matching then corresponds to assigning an orientation to each of the of $n$ edges of the quiver to satisfy the second condition. For the toric diagram, we assign $+1$ and $0$ contributions to the $x$ coordinate to the edges with right and left orientation, respectively. We obtain the toric diagram of $\mathbb{C}^{2}/\mathbb{Z}_{n}$, as expected.

 \subsection{Chiral fields and generalized perfect matchings}
    
\label{section_full_fields_pms}
    
Remarkably, the full field content of a perfect matching can be reconstructed from the knowledge of the chiral fields in it. 

Let us suppose that $\{\Phi_{i}^{(0)}\}$ is the set of chiral fields in a perfect matching $p$. The only terms in which fields of degree $m-1$ participate are the $J$-terms $\Phi^{(m-1)}_{a}J_{a}(\Phi_{i}^{(0)})$. The chiral fields in $p$ cover a subset of the $J$-terms so $p$ must contain all the $\Phi^{(m-1)}_{a}$ that appear in the rest of the terms. In order to satisfy the second condition, we include the conjugate of the remaining degree $m-1$ fields $\bar{\Phi}^{(1)}_{a}$, which determines all the field of degree $1$ in $p$. 

Continuing this process recursively, we can compute the full perfect matching. For every $1 \le k \le \lfloor\frac{m}{2}\rfloor$, the terms in which a field $\Phi^{(m-k)}_{a}$ appear have the form $\Phi^{(m-k)}_{a} P^{(k-1)}_{a}$, where $P^{(k-1)}$ is a polynomial of degree $k-1$ and hence only involves fields of degree $c\leq k-1$. The fields in $p$ of degree $c\leq k-1$ cover a subset of the superpotential terms and we must add the $\Phi^{(m-k)}_{a}$ appearing in the remaining ones. We will assume that every field appears in at least one term in the superpotential, so this unambiguously determines whether it is in $p$ or not. Once we establish the fields of degree $m-k$ in $p$, we must add conjugates of the remaining ones, which are the fields of degree $k$ in $p$. At the end of this process we will have computed all the fields in $p$ from the knowledge of the chiral fields in it.

\subsection{Perfect matchings for $\mathbb{C}^{m+2}$}

Let us illustrate the previous ideas, using perfect matchings to verify that the theories presented in \sref{section_Cm+2_permutohedra} indeed correspond to $\mathbb{C}^{m+2}$. 

In order to construct the perfect matchings, it is convenient to exploit the $SU(m+2)$ global symmetry. Picking a direction $\mu=1,\ldots, m+2$ breaks $SU(m+2)\to SU(m+1)$. Under this symmetry breaking, the 
quiver fields $\Phi^{(j-1;j)}$, $1\leq j \leq \floor{{m\over 2}}+1$, decompose as follows:\footnote{We have chosen to denote the degree of fields by $j-1$ for later convenience.} 
\beq
\Phi^{(j-1;j)} \to \Phi^{(j-1;j;\mu)} + \Phi^{(j-1;j;\cancel{\mu})} ~,
\eeq
where $\Phi^{(j-1;j;\mu)}$ and $\Phi^{(j-1;j;\cancel{\mu})}$ have $j-1$ and $j$ indices, respectively, and are explicitly given by:
\beq
\begin{array}{rcl}
          (\Phi^{(j-1;j;\mu)})^{\nu_{1}\cdots\nu_{j-1}} & = & (\Phi^{(j-1;j)})^{\mu\nu_{1}\cdots\nu_{j-1}} ~, \\[.25cm]
          (\Phi^{(j-1;j;\cancel{\mu})})^{\nu_{1}\cdots\nu_{j}} & = & (\Phi^{(j-1;j)})^{\nu_{1}\cdots\nu_{j}} ~ .
\end{array}
\eeq 
Making only the reduced $SU(m+1)$ symmetry manifest, the superpotential \eqref{potential_Cn} takes the form:
        \beq
        \begin{array}{rl}
          W \sim & \sum_{i+j+k = m+2}\Phi^{(j-1;j;\mu)}\Phi^{(k-1;k;\cancel{\mu})}\bar{\Phi}^{(m+1-j-k;m+2-j-k;\cancel{\mu})} \\[.25 cm]
          + & \sum_{i+j+k = m+2}\Phi^{(j-1;j;\cancel{\mu})}\Phi^{(k-1;k;\mu)}\bar{\Phi}^{(m+1-j-k;m+2-j-k;\cancel{\mu})} \\[.25 cm]
          + & \sum_{i+j+k = m+2}\Phi^{(j-1;j;\cancel{\mu})}\Phi^{(k-1;k;\cancel{\mu})}\bar{\Phi}^{(m+1-j-k;m+2-j-k;\mu)} ~ .
        \end{array}
        \eeq
It becomes clear that there are $m+2$ perfect matchings, one for each value of $\mu$. Furthermore, all $\Phi^{(j-1;j;\mu)}$ and $\bar{\Phi}^{(j-1;j;\mu)}$ form a perfect matching i.e.
\beq
p_{\mu} = \{\Phi^{(j-1;j;\mu)} , \bar{\Phi}^{(j-1;j;\mu)}| 1 \le j \le \frac{m}{2}+1\} ~.
\eeq
In particular, the chiral field content consists of a single chiral field:
\beq
p_{\mu}|_{\rm chiral}=\{ \Phi^{(0;1;\mu)} \} ~,
\eeq
i.e. there is a one-to-one correspondence between chiral fields and perfect matchings.

         We can choose the fundamental cycles of $\mathbb{T}^{m+1}$ such that 
        \begin{align}
          \ev{\Phi^{(0;1;\mu)},\gamma_{\alpha}} &= \delta_{\alpha,\mu}  ~,
        \end{align}
with $\alpha=1,\ldots, m+1$. In particular, this implies that $ \ev{\Phi^{(0;1;m+2)},\gamma_{\alpha}}=0$. The toric diagram therefore consists of the points:
        \begin{equation} 
        \begin{aligned}
          &v_0=(0,\ldots,0)~, \cr
          &v_1= (1,0,0,\ldots,0)~, \quad
           v_2=(0,1,0,\ldots,0)~, \quad \ldots~, \quad
           v_{m+1}=(0,0,\ldots,0, 1)~,
        \end{aligned}
        \end{equation}
which is indeed the toric diagram of $\mathbb{C}^{m+2}$.

\section{An infinite family: $F_{0}^{(m)}$}

\label{section_F0m}

\begin{figure}[ht]
	\centering
	\includegraphics[width=13cm]{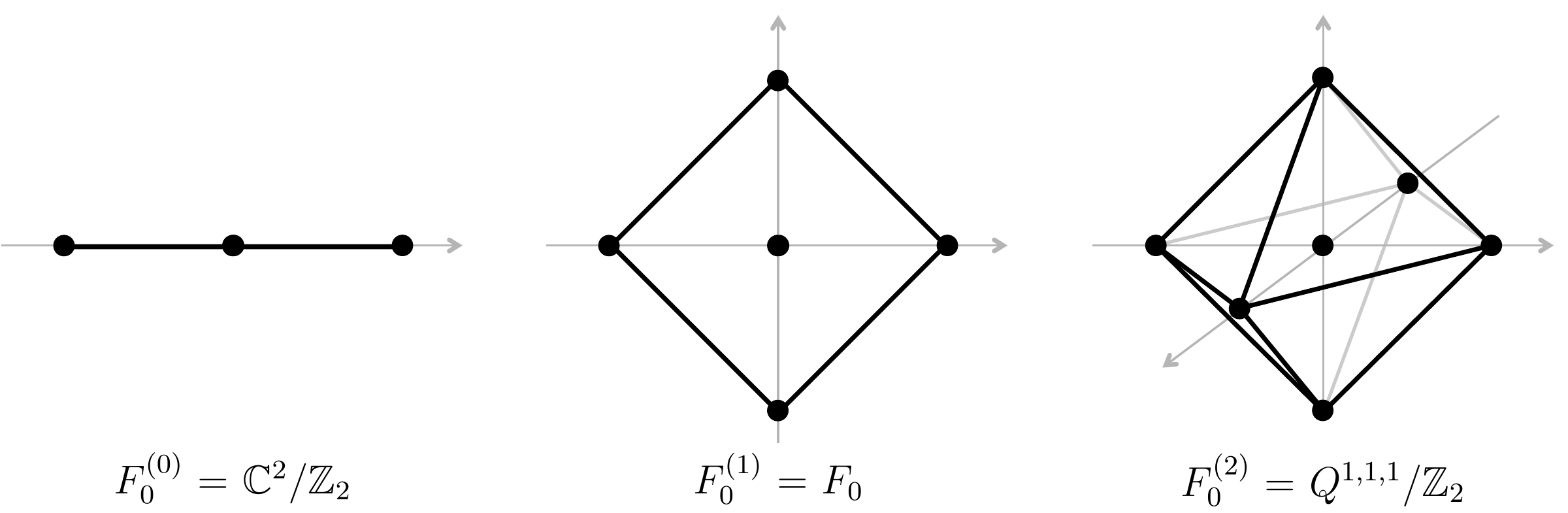}
\caption{Toric diagrams for $F_0^{(m)}$ with $m=0,1,2$.}
	\label{toric_diagrams_first_F0ms}
\end{figure}

Let us now consider another infinite family of toric geometries denoted $F_0^{(m)}$, which correspond to the affine cones over the $(\mathbb{P}^1)^{m+1}$. We will illustrate in detail how perfect matchings capture the moduli spaces of the corresponding quiver theories. The toric diagram for $F_0^{(m)}$ is the $(m+1)$-dimensional polytope consisting of the following points:
\begin{equation}
    \begin{array}{c}
    (0,\ldots,0) \\
    (\pm 1,0,\ldots, 0) \\
    \vdots \\
    (0,\ldots,0,\pm1)
     \end{array}
     \label{toric_F0m_general}
\end{equation}
$F_0^{(m)}$ has an $SU(2)^{m+1}$ isometry. The $\alpha^{th}$ $SU(2)$ factor acts on the toric diagram by permuting the two points with $\pm 1$ in the $\alpha^{th}$ coordinate and the origin is invariant under all $SU(2)$'s. For low $m$, this family contains some well-studied geometries: $ F_0^{(0)}  =  \mathbb{C}^2/\mathbb{Z}_2$, $F_0^{(1)} = F_0$ and $F_0^{(2)} = C(Q^{1,1,1}/\mathbb{Z}_2)$. \fref{toric_diagrams_first_F0ms} shows their toric diagrams.

This is an interesting class of geometries since it exhibits some of the main features of generic CY$_{m+2}$'s while being particularly tractable thanks to the large global symmetry.

\subsection{Quiver theories}

The quiver theories for the $F_0^{(m)}$ family were first introduced in \cite{Closset:2018axq}, where they were independently derived using a generalization of the orbifold reduction procedure \cite{Franco:2016fxm} and the topological B-model. More precisely, a toric phase was constructed for each of these geometries. Below we review them before computing their moduli spaces using perfect matchings.

\paragraph{Nodes.}   
The quiver has $\chi((\mathbb{P}^1)^{m+1})= 2^{m+1}$ nodes. We index every node $i$ by a binary vector $\vec{i}$ of length $m+1$. There is partial ordering relation $\succ$ on nodes defined as follows
        \begin{align}
            j \succ i \ \ \ \ \Leftrightarrow \ \ \ \  j_{\alpha} \ge i_{\alpha} \,\,\mbox{for all }\, \alpha=1,\cdots,m+1 ~.
        \end{align}

\begin{figure}[ht]
	\centering
	\includegraphics[width=14cm]{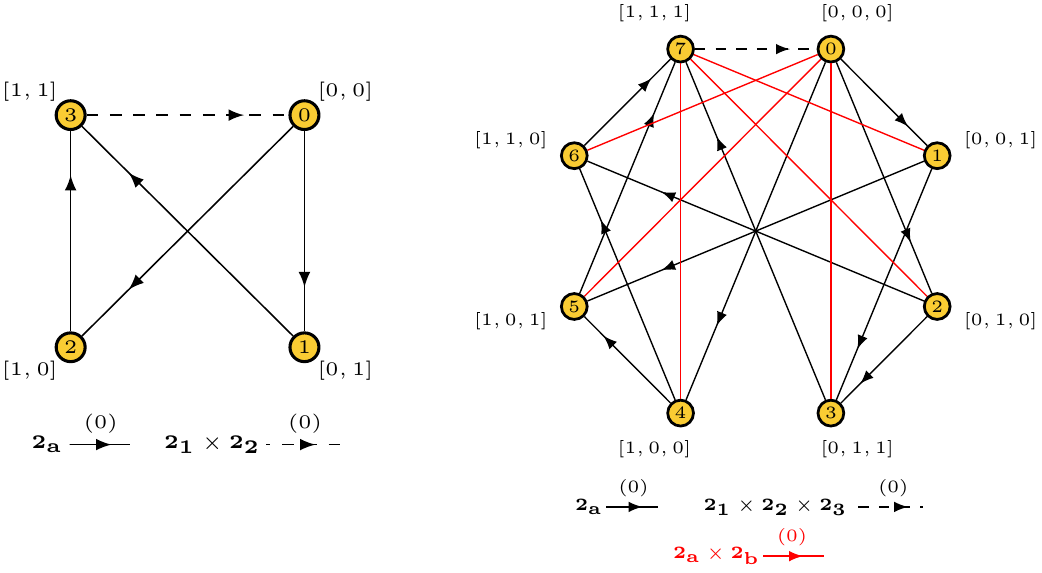}
\caption{Quiver diagrams for $F_0^{(1)}$ (left) and $F_0^{(2)}$ (right).}
	\label{quivers_F0}
\end{figure}

\paragraph{Arrows.}
Given two nodes $j$ and $i$ such that $j \ge i$, there is a multiplet $X_{ij}$ of arrows of degree $d_{ij}-1$ connecting them, where:
\beq
d_{ij} = \sum_{\alpha}(j_{\alpha} - i_{\alpha}) ~.    
\eeq
This multiplet contains $2^{d_{ij}}$ arrows which transform in the ${\bf{2}}_{1}^{j_{1}-i_{1}} \times \cdots \times {\bf 2}^{j_{m+1}-i_{m+1}}_{m+1}$ of the $SU(2)^{m+1}$ global symmetry, with the subindices labeling the different $SU(2)$ factors. \fref{quivers_F0} shows the quivers for $F_0^{(1)}$ and $F_0^{(2)}$, which correspond to phase 2 of $F_0$ \cite{Feng:2002zw} and  phase $L$ of $Q^{1,1,1}/\mathbb{Z}_2$ in the classification of \cite{Franco:2018qsc}, respectively.
        
\paragraph{Superpotential.}
            The superpotential is the most general cubic $SU(2)^{m+1}$ invariant of degree $m-1$. It is given by
            \begin{equation}
                W = \sum_{i}\sum_{j \succ i}\sum_{k \succ j}(-1)^{d_{ij}+m\,d_{ik}}X_{ij}X_{jk}\bar{X}_{ki} ~ ,
		\label{W_F0}
           \end{equation}
where we have suppressed the $SU(2)$ indices and the Levi-Civita tensors contracting them.

\paragraph{Periodic quiver.}
The periodic quiver for this family can be described straightforwardly. We take the fundamental domain of the torus to be the cube $[-1,1]^{m+1}$. This domain can be divided into $2^{m+1}$ quadrants, with each quadrant indexed by a vector of signs $\vec{q}$. Restricting to the quadrant with all $+$ signs, the quiver is the same as the quiver in the fundamental domain of $\mathbb{C}^{m+2}$ described above.

For any other quadrant $\vec{q}$, the quiver is obtained by a reflection with respect to the hyperplane: 
\beq
x_{\alpha} = 0 \qquad q_{\alpha} = - ~.
\eeq
It follows that, as for $\mathbb{C}^{m+2}$, all terms in the superpotential are cubic.

\subsection{Moduli space}

We are ready to explore how perfect matchings give rise to the $F_{0}^{(m)}$ moduli space. The discussion in this section significantly supersedes the preliminary analysis presented in \cite{Closset:2018axq}. In particular, we will provide explicit expressions for the perfect matchings and their multiplicity.

\subsubsection{Central perfect matchings and Dedekind numbers}

Let us first focus on the central point of the toric diagram \eref{toric_F0m_general}. Since this points is invariant under the global symmetry, the perfect matchings corresponding to it must contain complete representations of $SU(2)^{m+1}$. One such perfect matching is immediately evident from the superpotential \eref{W_F0}. It consists of all the arrows
\beq
p_0= \{\bar{X}_{ij}|i \succ j\} ~.
\label{p0}
\eeq
The chiral fields in this perfect matching are in $\bar{X}_{(1,\cdots,1),(0,\cdots,0)}$ which has dimension $2^{m+1}$ and transforms as $\mathbf{2}_{1} \times \cdots \times \mathbf{2}_{m+1}$. In the examples in \fref{quivers_F0}, these are the conjugate of the dotted arrows. 

This is not the only perfect matching associated to the central point. The multiplicity of perfect matchings corresponding to it rapidly grows with $m$. For example, it is known that for the phases under consideration the perfect matching multiplicity of the central point is $2$ for $F_{0}^{(0)}$, $5$ for $F_{0}^{(1)}$ and 19 for $F_{0}^{(2)}$.

The central perfect matchings can be elegantly classified in terms of Boolean functions. A Boolean function of $m+1$ variables is a function $f:\{0,1\}^{m+1} \to \{0,1\}$. For us, the domain of $f$ corresponds to the nodes of $F^{(0)}(m)$, which have a partial ordering $\succ$. A Boolean function $f$ is monotonically increasing if for any $i$ with $f(i) = 1$ we also have $f(j) = 1$ for all $j \succ i$.\footnote{Notice that this definition includes the constant functions $f=0$ and $f=1$.} 

Given a monotonically increasing Boolean function $f$, we define a collection of fields $\tilde{p}_{f}$ as follows:
\beq
\tilde{p}_{f} = \{X_{ij}|f(i) =0 \mbox{ and } f(j) = 1\} ~.
\label{pm_Boolean_1}
\eeq
Using $\tilde{p}_{f}$ we define:
\beq
p_{f} = \tilde{p} \cup \{\bar{X}_{ji}|X_{ij} \notin \tilde{p}_{f}\} ~.
\label{pm_Boolean_2}            
\eeq
In Appendix \ref{section_pms_F0} we show that the $p_{f}$ are indeed perfect matchings. In fact the previous definition also accounts for $p_0$. It is straightforward to see that following \eref{pm_Boolean_1} and \eref{pm_Boolean_2}, both constant functions $f=0$ and $f=1$ map to the same perfect matching $p_0$. It is clear that all these perfect matchings correspond to the central point of the toric diagram of $F_{0}^{(m)}$, since the $X_{ij}$'s (and their conjugates) represent full $SU(2)^{m+1}$ representations. The same conclusion is obtained by computing intersection numbers with the fundamental cycles.

The integer sequence $M(n)$ of the numbers of monotonically increasing Boolean functions is known as Dedekind numbers. The multiplicity of central perfect matchings is then:
\beq
            \# \, \mbox{central pm's of } F_{0}^{m} = M(m+1) - 1 ~,
\label{F0_multiplicities}
\eeq
where we have taken into account the fact that the two constant functions map to $p_0$. Dedekind numbers grow very quickly and only the values for $0 \le n \le 8$ are known explicitly \cite{MR1129608}. Combined with \eref{F0_multiplicities}, they give rise to the following multiplicities:
\beq
\begin{array}{|c|c|}
\hline 
\ m \ \ & \mbox{Multiplicity} \\ \hline \hline
0 & 2 \\ \hline
1 & 5 \\ \hline
2 & 19 \\ \hline
3 & 167 \\ \hline
4 & 7,580 \\ \hline
5 & 7,828,353 \\ \hline
6 & 2,414,682,040,997 \\ \hline
7 & 56,130,437,228,687,557,907,787 \\ \hline
\end{array}
\eeq
For $m\leq 2$, there is full agreement with the known results mentioned earlier. The multiplicities for $m>2$ are new predictions.

\subsubsection{Corner perfect matchings}

Next let us consider the corners of the toric diagram \eref{toric_F0m_general}, for which $x_{\mu} = \pm 1$ and all the other coordinates are zero. $SU(2)_{\mu}$ transforms these two points into one another, so picking one of them breaks $SU(2)_{\mu}$ down to $U(1)\times U(1)$. We need to consider how a representation $X_{ij}$ of $SU(2)^{m+1}$ splits under this reduced symmetry. There are two possibilities:
        \begin{itemize}
            \item
                $i_{\mu} = j_{\mu}$. In this case the original multiplet transforms trivially under $SU(2)_{\mu}$ and remains intact. The same is true for its conjugate.
            \item
                $j_{\mu} - i_{\mu} = 1$. In this case $X_{ij}$ splits into two multiplets: $X_{ij}^{+}$  and $X_{ij}^{-}$ both of which transform as
                \begin{align}
                    \mathbf{2}_{1}^{j_{1}-i_{1}}\times \cdots \times\mathbf{2}_{\mu-1}^{j_{\mu-1}-i_{\mu-1}}\times \mathbf{2}_{\mu+1}^{j_{\mu+1}-i_{\mu+1}}\times \cdots \times \mathbf{2}_{m+1}^{j_{m+1}-i_{m+1}}
                \end{align}
under the remaining $SU(2)^{m}$.

We again make all the quantum numbers explicit so that the conjugate of $X_{ij}^{+}$ is $\bar{X}_{ji}^{-}$.
        \end{itemize}
        The superpotential also splits into two parts
        \begin{align}
            W = W_{0} + W_{+-} ~.
        \end{align}
        $W_{0}$ consists of terms which contain no arrows charged under $SU(2)_{\mu}$. $W_{+-}$ consist of terms with two arrows charged under $SU(2)_{\mu}$, one barred and the other one unbarred. Under the reduced symmetry such a term splits as
        \begin{align}
            X_{ij}X_{jk}\bar{X}_{ki} \to X_{ij}^{+}X_{jk}\bar{X}^{-}_{ki} - X_{ij}^{-}X_{jk}\bar{X}^{+}_{ki} && j_{\mu} - i_{\mu} = 1 \nonumber \\
            X_{ij}X_{jk}\bar{X}_{ki} \to X_{ij}X_{jk}^{+}\bar{X}_{ki}^{-} - X_{ij}X_{jk}^{-}\bar{X}_{ki}^{+} && k_{\mu} - j_{\mu} = 1
        \end{align}
        With this, it is straightforward to verify that the following collection $p_{\mu}^{+}$ of arrows is a perfect matching:
        \begin{itemize}
            \item
                If $j_{\mu} - i_{\mu} = 1$, then $p_{\mu}^{+}$ contains $X_{ij}^{+}$ and the conjugate of $X_{ij}^{-}$, i.e. $\bar{X}_{ji}^{+}$. These arrows cover every term in $W_{+-}$ exactly once and do not cover any term in $W_{0}$.
            \item
                If $j_{\mu} - i_{\mu} = 0$, then $p_{\mu}^{-}$ contains $\bar{X}_{ji}$. These arrows cover every term in $W_{0}$ exactly once and do not cover any term in $W_{+-}$.
        \end{itemize}
        Above we have assumed that $j \succ i$, which is the condition for the existence of an arrow between $i$ and $j$. 

$p_{\mu}^{+}$ is the perfect matching which corresponds to $x_{\mu} = 1$. The chiral content of this perfect matching is then:
        \begin{align}
            p_{\mu}^{+} = \left\{X^{+}_{(a_{1},\cdots , a_{\mu-1} 0 , a_{\mu+1} , \cdots a_{m+1}),(a_{1},\cdots , a_{\mu-1} , 1 , a_{\mu+1} , \cdots a_{m+1})}\right\}\cup \left\{\bar{X}^{+}_{(1,\cdots,1),(0,\cdots,0)}\right\} ~.
        \end{align}
Similarly, the perfect matching corresponding to $x_{\mu} = -1$, which we denote $p_{\mu}^{-}$, is the following collection of arrows:
        \begin{itemize}
            \item
                If $j_{\mu} - i_{\mu} = 1$, then $p_{\mu}^{-}$ contains $X_{ij}^{-}$ and the conjugate of $X_{ij}^{+}$ i.e $\bar{X}_{ji}^{-}$.
            \item
                If $j_{\mu} - i_{\mu} = 0$, then $p_{\mu}^{-}$ contains $\bar{X}_{ji}$. 
        \end{itemize}
        The chiral content of this perfect matching is:
        \begin{align}
            p_{\mu}^{-} = \left\{X^{-}_{(a_{1},\cdots , a_{\mu-1} 0 , a_{\mu+1} , \cdots a_{m+1}),(a_{1},\cdots , a_{\mu-1} , 1 , a_{\mu+1} , \cdots a_{m+1})}\right\}\cup \left\{\bar{X}^{-}_{(1,\cdots,1),(0,\cdots,0)}\right\} ~.
        \end{align}

\section{A simplified algorithm for finding perfect matchings}

\label{section_simplified_algorithm}

We have seen that perfect matchings provide a simple combinatorial approach for calculating the moduli space of $m$-dimers and illustrated their applicability with two infinite families of theories. While this represents a significant simplification with respect to alternative methods, in this section we introduce a considerably more efficient algorithm for computing perfect matchings, one that does not rely on the direct application of their definition. This can be regarded as a generalization to arbitrary $m$ of the elegant approach based on the Kasteleyn matrix for brane tilings. In order to set up the stage for the new method, we first revisit the Kasteleyn matrix from a new perspective and also consider the counting of brick matchings for brane brick models.

\subsection{Warm up: toric CY 3-folds and brane tilings}

\label{section_Kasteleyn}

As a warm up, we first consider the familiar case of brane tilings and how their perfect matchings can be determined using the Kasteleyn matrix.

\subsubsection{The Kasteleyn matrix revisited}

\label{section_Kasteleyn_revisited}

The superpotential of a brane tiling consists of an equal number of positive terms  $W^{+}_{a}$  (white nodes) and negative terms $W^{-}_{b}$ (black nodes). The perfect matchings can be neatly packaged into the Newton polynomial:
\beq
P(x,y) = \det(K) ~,
\label{P_det_K}
\eeq
where the Kasteleyn matrix $K$ is defined as follows:
\beq
K_{ab} = \sum_{i \, \in  \, a,b} X_i \, x^{\vev{X_{i},\gamma_{x}}} y^{\vev{X_{i},\gamma_{y}}} ~.
\eeq
The index $i$ labels the edges in the brane tiling. Every entry in $K$ is thus given by the sum over all edges connecting the corresponding pair of nodes (equivalently, the sum over all chiral fields participating in the associated pair of superpotential terms). Furthermore, every edge is weighted by a monomial in $x$ and $y$ that encodes its intersection numbers with 
$\gamma_{x}$ and $\gamma_{y}$, the fundamental cycles of $\mathbb{T}^2$.\footnote{Minor variations of this definition exist, depending on whether individual edges of the tiling are labeled (as in our expression) and additional signs are included.}

For later generalizations, it is convenient to rewrite \eref{P_det_K} as Grassmann integral. To every $W^{+}_{a}$ and $W^{-}_{b}$ we associate Grassmann variables  $\theta_{a}^{+}$ and $\theta^{-}_{b}$, respectively. Then we get:
\beq
P(x,y) = \int\prod_{a}\ud\theta^{+}_{a}\ud\theta^{-}_{a}\exp\left(\sum_{i}\Theta(X_{i})X_{i} \, x^{\vev{X_{i},\gamma_{x}}}y^{\vev{X_{i},\gamma_{y}}}\right) \label{npgon},
\eeq
where the function $\Theta(X_{i})$ is the product of the Grassmann variables associated to the pair of superpotential terms in which $X_{i}$ occurs:
\beq
\Theta(X_{i}) = \theta^{+}_{a}\theta^{-}_{b} \ \ \ \ \ \ \ \mbox{for } X_i \in W^{+}_{a}, W^{-}_{b} ~.
\eeq

\subsubsection{Permanent vs determinant}

The coefficient of the $x^m y^n$ term in $P(x,y)$ defined as above is the sum (up to signs) of the perfect matchings, expressed as the products of the fields in them, corresponding to the point with coordinates $(m,n)$ in the toric diagram. The signs in the determinant correspond to the anticommutativity of Grassmann variables. Although their squaring to zero is essential for \eqref{npgon}, anticommutativity is not. We can alternatively use commuting variables, i.e. we define these variables by
        \begin{align}
            \theta_{a}^{2} &= 0 \nonumber \\
            \theta_{a}\theta_{b} &= \theta_{b}\theta_{a} 
        \end{align}
and we define the integration in the same way as for normal Grassmann variables
        \begin{align}    
            \int \ud\theta_{a}\,1 &= 0 \nonumber \\
            \int \ud\theta_{a}\,\theta_{a} &= 1  
        \end{align} 
Here we note that these properties follow immediately if we consider each of these variables to be the product of two independent Grassmann variables.

Computing \eqref{npgon} with this definition of $\theta^{\pm}_{i}$ will give us the permanent of the Kasteleyn matrix. In all the discussions that follow, we can either regard Grassmann variables in the usual sense or as this modification.

\subsection{CY 4-folds and brane brick models}

Finding perfect matchings of a brane tiling using the Kasteleyn matrix relies crucially on the fact that every chiral field participates in two superpotential terms with different signs.  But for brane brick models, different chiral fields can take part in different numbers of $J$- and $E$-terms (see e.g. \cite{Franco:2015tya} for explicit examples). Therefore, we do not expect that the Newton polytope can be expressed as a determinant. However, its formulation as an integral over some auxiliary Grassmann variables is more amenable to generalization.  In this section we will investigate such extensions, progressively simplifying them, before moving to general $m$-dimers.

\subsubsection{First approach: Grassmann variables for plaquettes}

Recall the combinatorial definition of perfect matchings for brane brick model, i.e. brick matchings, given in \sref{section_BBMs}. A brick matching $p$ is a collection of chiral and Fermi fields such that:
    \begin{itemize}
        \item 
            For every Fermi field $\Lambda_{a}$, $p$ contains exactly either $\Lambda_{a}$ or $\bar{\Lambda}_{a}$.
        \item
            If $p$ contains $\Lambda_{a}$, it contains exactly one chiral field in each of $E_a^+$ and $E_a^-$.
        \item
            If $p$ contains $\bar{\Lambda}_{a}$, it contains exactly one chiral field in each of $J_a^+$ and $J_a^-$.
    \end{itemize}

Given this definition, we can write an expression analogous to \eqref{npgon}. To do so, we associate a Grassmann variable to every $J$- and $E$-term. There are four variables per Fermi field: $\theta_{a}^{\pm}$ from $J^{\pm}$ and $\bar{\theta}^{\pm}_{a}$ from $E_{a}^{\pm}$. For every chiral $X_{i}$, we define $\Theta(X_{i})$ as the following product:
\begin{itemize}
\item Every $J^{+}_{a}$ or $J^{-}_{a}$ term containing $X_{i}$ contributes a $\theta_{a}^{+}\bar{\Lambda}_{a}$ or $\theta_{a}^{-}\bar{\Lambda}_{a}$ factor, respectively.
\item Every $E^{+}_{a}$ or $E^{-}_{a}$ term containing $X_{i}$ contributes a $\bar{\theta}_{a}^{+}\Lambda_{a}$ or $\bar{\theta}_{a}^{-}\Lambda_{a}$ factor, respectively.
\end{itemize}
From a brane brick model perspective, the Grassmann variables in $\Theta(X_{i})$ are simply those attached to edges shared by $X_{i}$ and Fermi fields.

The Newton polynomial for a brane brick model is then given by
\begin{align}
P(x,y,z) = \int \prod_{a}(\ud\theta^{+}_{a}\ud\theta^{-}_{a} + \ud\bar{\theta}^{+}_{a}\ud\bar{\theta}^{-}_{a})\exp\left(\sum_{i}\Theta(X_{i})\, X_{i}\, x^{\vev{X_{i},\gamma_{x}}}y^{\vev{X_{i},\gamma_{y}}}z^{\vev{X_{i},\gamma_{z}}}\right) \label{mdint} ~.
\end{align}

\subsubsection{Second approach: Grassmann variables for chiral cycles}

\label{section_pms_chiral_cycles}

Although \eqref{mdint} provides an algebraic expression for the Newton polynomial, it is considerably hard to work with. Instead of one top-level integral as in the case of dimer models, it gives rise to a collection of mid-dimensional integrals whose number grows rapidly with the number of Fermis. 

In order to remedy this, it is convenient to introduce an equivalent definition of brick matchings. From now on, we will focus on their chiral field content. The reason for doing this is twofold: the Fermi content of a brick matching is fixed by the chiral fields in it and the toric diagram only depends on the chiral fields. It is straightforward to reintroduce the Fermis, if necessary.

Let us consider the $J$- and $E$-terms associated to a Fermi field $\Lambda_{a}$. The product 
\begin{align}
J_{a}E_{a} = J^{+}_{a}E^{+}_{a} - J^{+}_{a}J^{-}_{a} - J^{-}_{a}E^{+}_{a} + J^{-}_{a}E^{-}_{a}
\label{chiral_cycles_BBM}
\end{align}
is a sum of four {\it chiral cycles}. We can alternatively define the chiral content of a brick matching as a collection of chiral fields that contains exactly one field from each of these chiral cycles for every Fermi field. It is easy to see that a brick matching as defined above will have two (not necessarily distinct) chiral fields from the $J$- and $E$-terms of a given Fermi field $\Lambda_{a}$, and either both of them belong to $J_{a}$ or both belong to $E_{a}$. Hence, it covers either both $J$-terms and we add $\bar{\Lambda}_{a}$ to it or it covers only $E$-terms and we add $\Lambda_{a}$. With this completion with Fermi fields, this definition is clearly equivalent to the previous one.

Now it is easy to give an expression for the Newton polytope as a top level integral over auxiliary Grassmann variables. This time we assign $\theta^{ss^{\prime}}_{a}$ to the chiral cycle $J^{s}_{a}E^{s^{\prime}}_{a}$. Again, there are four variables per Fermi. The Newton polynomial becomes
\begin{align}
P(x,y,z) = \int \prod_{a}\ud\theta^{++}_{a}\ud\theta^{+-}_{a}\ud\theta^{-+}_{a}\ud\theta^{--}_{a}\exp\left(\sum_{i}\Theta(X_{i})\, X_{i}\, x^{\vev{X_{i},\gamma_{x}}}y^{\vev{X_{i},\gamma_{y}}}z^{\vev{X_{i},\gamma_{z}}}\right) \label{tlint} ~,
\end{align}
where $\Theta(X_{i})$ is now defined as the product of the Grassmann variables for all the chiral cycles containing $X_i$.

\subsubsection{Final approach: further simplification using the trace condition}

\label{trace_condtion_grassmann_formula}

The previous expression admits a further simplification. The so-called trace condition,
\beq 
\sum_{a} J_{a}E_{a}= 0 ~,
\eeq
is required by $2d$ $(0,2)$ supersymmetry and is equivalent to the vanishing of the Kontsevich bracket \eref{superpotential_Kontsevich} in the $m=2$ case \cite{Franco:2017lpa}. Due to the trace condition, it is clear that applying \eref{chiral_cycles_BBM} to all Fermis, every chiral cycle will be generated twice (with opposite signs). Since we are just interested in counting different chiral cycles, we can reduce the number of Grassmann integrations by half, e.g. by picking the chiral cycles that occur in this expansion with a positive sign. Hence, if we assign variables $\theta^{+}_{a}$ to the cycle $J_{a}^{+}E_{a}^{+}$ and $\theta^{-}_{a}$ to the cycle $J^{-}_{a}E^{-}_{a}$ the Newton polynomial can be computed as
\begin{align}
P(x,y,z) = \int \prod_{a}\ud\theta^{+}_{a}\ud\theta^{-}_{a}\exp\left(\sum_{i}\Theta(X_{i})\, X_{i} \, x^{\vev{X_{i},\gamma_{x}}}y^{\vev{X_{i},\gamma_{y}}}z^{\vev{X_{i},\gamma_{z}}}\right) ~,\label{bmfi}
\end{align}
where $\Theta(X_{i})$ now contains a $\theta^{+}_{a}$ factor iff $J^{+}_{a}$ or $E^{+}_{a}$ contains $X_{i}$ and  a $\theta^{-}_{a}$ factor iff $J^{-}$ or $E^{-}$ contains $X_{i}$.

\subsection{An algorithm for general $m$}

\label{section_chiral_cycles_general_m}
    
Starting from the definition of perfect matchings for general $m$-dimers given in \sref{section_generalized_pms} we can immediately write and expression that computes them, analogous to \eqref{mdint}. While correct, such formula would have the same drawbacks we mentioned earlier:
\begin{itemize}
\item Since for every field $\Phi$, either $\Phi$ or $\Bar{\Phi}$ is in the perfect matching, the expression would involve numerous factors in the measure and hence several non top-dimensional integrals.

\item It is desirable to focus on chiral fields only, since they are sufficient for reconstructing the full perfect matchings and for determining the moduli space. This is a considerable simplification since, in general, the number of chiral fields is significantly lower than the total number of fields. For example, the  $\mathbb{C}^{m+2}$ quiver contains $2^{m+1}-1$ fields but only $m+2$ of them are chiral.
\end{itemize}

\paragraph{Chiral cycles.}

Due to these reasons, an expression as a top-dimensional integral defined in terms of chiral fields only is very attractive, both conceptually and computationally. This can be achieved by extending the concept of {\it chiral cycles} to general $m$. Chiral cycles are oriented cycles in the quiver which only contain chiral fields and are defined as follows:
\begin{itemize}
\item The superpotential can be written as:\footnote{In the rest of this section, we do not care about signs or numerical factors and regard sums simply as collections of cycles. In particular, if a given cycle appears twice with opposite signs, we do not cancel the two contributions but keep a single term indicating the presence of the cycle.}
\begin{align}
W = \tilde{W} + \sum_{a}\Phi^{(m-1)}_{a}J_{a} ~,
\end{align}
where $\tilde{W}$ and $J_{a}$ do not involve fields of degree $m-1$. For a term $\tilde{W}_{r}\in \tilde{W}$, let us define $\tilde{W}_{r}^{(1)}$ by:
\begin{align}
\tilde{W}_{r}^{(1)} = \eval{\tilde{W}_{r}}_{\bar{\Phi}^{(1)}_{a} = J_{a}} ~,     
\label{replacement_W_tilde_1}
\end{align}
i.e. we evaluate every $\bar{\Phi}^{(1)}_{a}$ at the corresponding $J$-term $J_{a}$. 

We denote $W^{(1)}$ the result of replacing all the terms $\tilde{W}_{r}$ in the superpotential according to \eref{replacement_W_tilde_1}. $W^{(1)}$ is a sum of cycles that neither contain fields of degree $m-1$ nor their conjugates, i.e. fields of degree $1$. Note that this process reduces the total degree of a term by the number of $\bar{\Phi}^{(1)}_{a}$ in it. As for brane brick models, due to the vanishing Kontsevich bracket condition, this process generates multiple copies of the same cycles (with signs). As previously mentioned, we do not cancel such contributions and count every cycle once. We apply the same procedure in the steps that follow. 

\item We continue this process recursively, defining $W^{(k+1)}$ for $1 \le k \le \lfloor\frac{m}{2}\rfloor$ as follows. Suppose $W^{(k)}$ is a sum of cycles which do not contain fields of degree $1 \le i \le k$ or their conjugates. Then $W^{(k)}$ can be written as:
\begin{align}
W^{(k)} = \tilde{W}^{(k)} + \sum_{a}\Phi^{(m-k-1)}_{a}J^{(k)}_{a} \label{wjk} ~,
\end{align}
where $W^{(k)}$ and $J^{(k)}_{a}$ do not involve fields of degree $m-k-1$. This form follows immediately from the fact that every term in $W$ has degree $m-1$ and hence the terms in $W^{(k)}$ have degree less than $m-1$.

For a term $\tilde{W}^{(k)}_{r}$ of $\tilde{W}^{(k)}$ we define $\tilde{W}_{r}^{(k+1)}$ as:
\begin{align}
\tilde{W}_{r}^{(k+1)} = \eval{\tilde{W}_{r}^{(k)}}_{\bar{\Phi}^{(m-k-1)}_{a} = J^{(k)}_{a}} ~.
\end{align}
We obtain a collection of cycles that do not contain fields of degree $1 \le i \le k+1$. We call $W^{(k+1)}$ the sum of the independent cycles obtained at this step. 

\item This process terminates with $W^{(k_{max})}$, where $k_{max} = \lfloor\frac{m}{2}\rfloor$, which is a collection of cycles $W^{(k_{max})}_{r}$ consisting entirely of chiral fields. These are the chiral cycles we are interested in.
\end{itemize}

\subsubsection{From chiral cycles to perfect matchings and the toric diagram}

\label{section_pm_chiral_cycles_general}

It is straightforward to verify that every perfect matching $p$ contains exactly one field from each term of $W^{(k)}$, for $0 \le k \le k_{max}$. In particular, this is true for $W^{(k_{max})}$, i.e. a perfect matching contains exactly one chiral field from every chiral cycle. Moreover, such a collection of chiral fields can be uniquely completed into a perfect matching using the process described earlier. This provides an alternative definition of perfect matchings based on chiral cycles.

We can now write a simple Grassmann integral that efficiently computes the perfect matchings and their positions in the toric diagram. It is given by:
\begin{align}
P(x_{\mu}) = \int \prod_{r}\theta_{r} \exp(\sum_{i}\Theta(\Phi^{(0)}_{i})\, \Phi^{(0)}_{i}\, \prod_{\mu}x_{\mu}^{\vev{\Phi^{(0)}_{i},\gamma_{\mu}}}) ~,
\label{general_Grassmann_integral}
\end{align}
where $\theta_{r}$ is the Grassmann variable associated to the chiral cycle $W^{(k_{max})}_{r}$ and $\Theta(\Phi^{(0)}_{i})$ is the product of the Grassmann variables for the chiral cycles that contain $\Phi^{(0)}_{i}$.

\subsection{Chiral cycles for small $m$}
    
For small $m$ we can easily enumerate all possible types of superpotential terms. As a result, it is also possible to classify the different types of chiral cycles. Below we present this classification for $m\leq 4$.

\paragraph{m=0.} 
There is no superpotential so there are no chiral cycles.

\medskip

\paragraph{m=1.} 
In this case the superpotential is a holomorphic function of the chiral fields so every term in the superpotential is a chiral cycle. As explained in \sref{section_Kasteleyn}, assigning a Grassmann variable to each of them, \eref{general_Grassmann_integral} is a Gaussian integral which evaluates to the permanent of the Kasteleyn matrix.

\medskip

\paragraph{m=2.} 
As discussed in \sref{section_BBMs} the superpotential in this case has the general form
\begin{align}
W = \sum_{a}\Lambda_{a}(J_{a}^{+}(X_{i}) - J_{a}^{-}(X_{i})) + \bar{\Lambda}_{a}(E_{a}^{+}(X_{i}) - E_{a}^{-}(X_{i}) ) ~.
\end{align} 
Hence, every Fermi field gives rise to four chiral cycles cycles
         \begin{align*}
             J_{a}^{+}E_{a}^{+} , J_{a}^{-}E_{a}^{-} , J_{a}^{+}E_{a}^{-} , J_{a}^{-}E_{a}^{-} ~.
         \end{align*}
As we mentioned in \sref{trace_condtion_grassmann_formula}, chiral cycles are generated multiple times due to the trace condition. We can obtain the independent chiral cycles by restricting to $J_{a}^{+}E_{a}^{+}$ and $J_{a}^{-}E_{a}^{-}$ for every $\Lambda_{a}$.

\medskip

\paragraph{m=3.} 
For $m=3$, there are two types of fields: chiral fields $X_{i}$ of degree $0$ and Fermi fields $\Lambda_{a}$, which by convention we take of degree $2$ \cite{Franco:2016tcm,Franco:2017lpa}.

The most general superpotential obeying the degree constraint has the form:
\beq
             W = \sum_{a}\Lambda_{a}J_{a}(X_{i}) + \sum_{a,b}\bar{\Lambda}_{a}\bar{\Lambda}_{b}H_{ab}(X_{i}) ~,
\eeq
where $J_{a}$ and $H_{ab}$ are holomorphic functions of chiral fields. Vanishing of $\{W,W\}$ means that for every $\Lambda_{a}$:
\beq
\sum_{b} H_{ab}J_{b} = 0 ~.
\label{m3_Kontsevich_bracket_constraint}
\eeq

Computing chiral cycles with the procedure introduced in \sref{section_chiral_cycles_general_m}, we get:
\beq
 W^{(1)} = \sum_{ab} H_{ab}J_{a}J_{b} ~.
\eeq
As mentioned earlier, throughout this section we regard sums simply as collections of cycles, without caring about numerical factors or implementing cancellations. These cycles are composed entirely of chiral fields and are the chiral cycles.

\paragraph{m=4.}
For $m=4$, the most general superpotential compatible with the degree condition takes the form:
\beq
W = \sum_{a}\bar{\Lambda} J_{a}(X_{i}) + \sum_{a,\alpha}\left[\chi_{\alpha}\Lambda_{a}H_{\alpha a}(X_{i}) + \bar{\chi}_{\alpha}\Lambda_{a}\tilde{H}_{\alpha a}(X_{i})\right] + \frac{1}{6}\sum_{a,b,c}\Lambda_{a} \Lambda_{b}\Lambda_{c}K_{abc}(X_{i}) ~,  
\eeq
where $X_{i}$, $\Lambda_a$ and $\chi_{\alpha}$ have degree 0, 1 and 2, respectively. $J_{a},H_{\alpha a}, \tilde{H}_{\alpha a}$ and $K_{abc}$ are holomorphic functions of chiral fields and $K_{abc}$ is antisymmetric under the exchange of any two indices. As for any even $m$, there is a symmetry under the exchange of $\chi_{\alpha} \leftrightarrow \bar{\chi}_{\alpha}$ and the simultaneous exchange of $H_{\mu \alpha}\leftrightarrow \tilde{H}_{\mu \alpha}$.

The vanishing of the Kontsevich bracket $\{W,W\}$ translates into the following conditions:
\beq
\begin{array}{l}
            \sum_{a}H_{\alpha a}J_{a} = \sum_{a}\tilde{H}_{\alpha a}J_{a} = 0 \\[.25cm]
            \sum_{\alpha}\left[H_{\alpha a}\tilde{H}_{\alpha b} - H_{\alpha b}\tilde{H}_{\alpha a}\right] + 2\sum_{c}K_{abc}J_{c} = 0
\end{array}            
\eeq

Let us now construct the chiral cycles from the superpotential, starting from the terms $W_{\alpha a} = \chi_{\alpha}\Lambda_{a}H_{\alpha a} + \bar{\chi}_{\alpha}\Lambda_{a}\tilde{H}_{\alpha a}$. We see that:
\beq
W^{(1)}_{\alpha a} = \chi_{\alpha}J_{a}H_{\alpha a} + \bar{\chi}_{\alpha}J_{a}\tilde{H}_{\alpha a} ~.
\eeq
These cycles still contain $\chi_{\alpha}$ and $\bar{\chi}_{\alpha}$, so we need to iterate the process once more in order to replace them with chiral fields. We obtain:
\beq
W^{(2)}_{\alpha a} = \sum_{b}J_{b}\tilde{H}_{\alpha b}J_{a}H_{\alpha a} ~.
 \label{m4_Wa}
\eeq

Similarly, let us consider the terms $W_{abc} = \Lambda_{a} \Lambda_{b}\Lambda_{c}K_{abc}$. They give rise to the additional chiral cycles:   
\beq
W^{(1)}_{abc} = J_{a}J_{b}J_{c}K_{abc} ~ .
 \label{m4_Wabc}            
\eeq    

Combining \eref{m4_Wa} and \eref{m4_Wabc} we obtain all the chiral cycles for these theories, which are of two kinds:
        \begin{align}
            C_{ab\alpha} &\sim J_{a}H_{\alpha a}J_{b}\tilde{H}_{\alpha b} \nonumber \\
            C_{abc}   &\sim J_{a}J_{b}J_{c}K_{abc}   
        \end{align} 
As usual, every cycle is generated multiple times due to the relations coming from the vanishing of $\{W,W\}$.

\section{Chiral cycles and perfect matchings for $Y^{1,0}(\mathbb{P}^{m})$}

\label{section_Y10}
    
We now illustrate the new algorithm introduced in the previous section with another infinite family of geometries, denoted $Y^{1,0}(\mathbb{P}^{m})$ \cite{Closset:2018axq}. The perfect matchings for this family of theories were already presented without a derivation in \cite{Closset:2018axq}. We now explain in detail how they are systematically determined using chiral cycles combined with Grassmann integrals. 

In order to make our presentation self-contained, we begin with a brief description of this family. The toric diagram for this family of singularities is given by:
\beq
v_0=(0,\ldots,0)~,
\qquad
\begin{array}{l}
v_1= (1,0,0,\ldots,0)~, \\
 v_2=(0,1,0,\ldots,0)~, \\
 \vdots \\
 v_{m+1}=(0,0,\ldots,0, 1)~, 
\end{array} 
\qquad
v_{m+2}=(1,1, \cdots, 1, 1)~.
 \label{eq_toric_Y10}
 \eeq 
The geometries have an $SU(m+1)$ isometry that translates into an $SU(m+1)$ global symmetry in the corresponding graded quivers. This symmetry acts by permuting the points $v_1,\ldots,v_{m+1}$. 

Let us now review the quiver theories for these geometries, which were first derived in \cite{Closset:2018axq} by a combination of the $3d$ printing algorithm of \cite{Franco:2018qsc} with partial resolution and, independently, using the topological B-model. The quiver has $m+1$ nodes which we will label by $0, \cdots , m$. The arrows are:
    \begin{align}
             X_{m,0}                         & : &m &\xrightarrow[(0)]{\phantom{aabbccd}1\phantom{aabbccd}}0 \nonumber \\
             X_{i+1,i}                       & : &i+1&\xrightarrow[(0)]{\phantom{aabbccd}1\phantom{aabbccd}}i        && 0 \le i \le m-1 \nonumber\\                   
             \Lambda_{i,i+k}^{(k-1;k)}       & : &i&\xrightarrow[(k-1)]{\phantom{aabb\,\,}\binom{m+1}{k}\phantom{bbcc\,\,}} i+k && 0 \le i \le m-1 ; 1 \le k \le m-i \nonumber\\
             \Gamma_{i,i+k}^{(k+1;k+1)}      & : &  i&\xrightarrow[(k+1)]{\phantom{aabb\, \, }\binom{m+1}{k+1}\phantom{bbcc\, \, }} i+k && 1 \le i \le m-1 ; 0 \le k \le m-i 
            \label{field_content_Y10m}
    \end{align}
The subscripts, taken ${\rm mod}(m+1)$, indicate the nodes connected by the arrows. $X_{m,0}$ and $X_{i+1,i}$ are chirals and singlets under $SU(m+1)$. For the rest of the arrows, we use a notation consisting of two superindices. The first integer is the degree of the field. The second integer $j$ indicates that the arrows transform in the $j$-index totally antisymmetric representation of $SU(m+1)$. In \eref{field_content_Y10m}, the numbers over the arrows are the dimensions of the $SU(m+1)$ representations and the numbers below are the degrees. We refer the reader to \cite{Closset:2018axq} for figures showing these intricate quivers up to $m=6$.

Let us consider the superpotential. All the terms in it are invariant under the $SU(m+1)$ global symmetry. The products of arrows we will write are explicitly given by
                \begin{align}
                    (A_{1}^{(c_{1};k_{1})}\cdots A_{n}^{(c_{n};k_{n})})^{\alpha_{k+1}\cdots\alpha_{m+1}} \equiv \frac{1}{\prod_{i}k_{i}!}\epsilon^{\alpha_{1}\cdots\alpha_{m+1}}A_{1;\alpha_{1}\cdots \alpha_{k_{1}}}^{(c_{1};k_{1})}\cdots A_{n;\alpha_{k-k_{n}+1}\cdots \alpha_{k}}^{(d_{n};l_{n})} \, ,
                    \label{convention_products}
                \end{align} 
                where $k = \sum_{i}k_{i}$ and the $\alpha_\mu$'s are fundamental $SU(m+1)$ indices. With this convention, any term with a total of $m+1$ indices is an $SU(m+1)$ invariant.

The superpotential consists of cubic terms $W_{3}$ and quartic terms $W_{4}$. The cubic terms are:
            \begin{align}
            W_{3} &= 
                \sum_{i=2}^{m}\sum_{k = 0}^{i-1}s_{1}(i,k)X_{i,i-1}\bar{\Gamma}_{i-1,i-1-k}^{(m-k-1;m-k)}\Lambda_{i-1-k,i}^{(k;k+1)} \nonumber\\
            & + \sum_{i=2}^{m}\sum_{k = 1}^{m-i}s_{2}(i,k)X_{i,i-1}\Lambda_{i-1,i-1+k}^{(k-1;k)}\bar{\Gamma}_{i-1+k,i}^{(m-k;m+1-k)} \nonumber\\
            & + \sum_{i=1}^{m-1}\sum_{k=1}^{i-1}\sum_{j=k}^{m-1-i}s_{3}(i,j,k)\Lambda_{i-k,i}^{(k-1;k)}\bar{\Gamma}_{i,i-j}^{(m-j-1;m-j)}\Gamma_{i-j,i-k}^{(j-k+1,j-k+1)} \nonumber\\
        \end{align}           
            
             \begin{align}           
            & + \sum_{i=1}^{m-1}\sum_{k=1}^{i-1}\sum_{j=0}^{m-i-1}s_{4}(i,j,k)\Lambda_{i-k,i}^{(k-1;k)}\Gamma_{i,i+j}^{(j+1;j+1)}\bar{\Gamma}_{i+j,i-k}^{(m-j-k-1;m-j-k)} \nonumber\\
            & + \sum_{i=1}^{m}\sum_{k=1}^{i-1}\sum_{j=1}^{m-i}s_{5}(i,j,k)\Lambda_{i-k,i}^{(k-1;k)}\Lambda_{i,i+j}^{(j-1;j)}\bar{\Lambda}_{i+j,i-k}^{(m+1-j-k;m+1-j-k)}
        \end{align}
and the quartic terms are:
        \begin{align}
           W_{4} =  &\sum_{k=1}^{m}s_{6}(k)X_{k,k-1}\Lambda_{k-1,m}^{(m-k,m-k+1)}X_{m,0}\Lambda_{0,k}^{(k-1;k)} \nonumber \\
                   +  & \sum_{k=1}^{m-1}\sum_{j=0}^{m-1-k}s_{7}(j,k)\Gamma_{k,k+j}^{(j+1;j+1)}\Lambda_{k+j,m}^{(m-k-j-1;m-k-j)}X_{m,0}\Lambda_{0,k}^{(k-1;k)}
        \end{align}
where $s_1,\cdots,s_7$ are signs which can be fixed by imposing $\{W,W\}=0$.

\bigskip
        
\subsection{Chiral cycles and the moduli space}

Knowing the superpotential, we are ready to find the chiral cycles for this family. Since to get them we substitute fields in the superpotential terms by polynomials with the same quantum numbers, chiral cycles will arise in $SU(m+1)$ invariant combinations. The chiral fields in the quiver are $X_{m,0}$, $X_{i+1,i}$ and $\Lambda_{i,i+1}^{(0;1)}$. In terms of them, the chiral cycles are:
\beq
\begin{array}{c}
            \Lambda_{0,1}^{(0;1)}X_{1,0}\Lambda_{0,1}^{(0;1)}\Lambda^{(0;1)}_{1,2} \cdots \Lambda^{(0;1)}_{m-1,m}X_{m,0} \\[.25cm]
            + \Lambda_{0,1}^{(0;1)}\Lambda^{(0;1)}_{1,2}X_{2,1}\Lambda_{1,2}^{(0;1)}\cdots\Lambda_{m-1,m}^{(0;1)}X_{m,0} \\[.25cm]
            + \cdots + \Lambda_{0,1}^{(0;1)}\Lambda^{(0;1)}_{1,2}\cdots \Lambda^{(0;1)}_{m-1,m}X_{m,m-1}\Lambda_{m-1,m}^{(0;1)}X_{m,0} 
\label{chiral_cycles_Y10}
\end{array}
\eeq      
Notice that despite every term containing a product of the form $\Lambda^{(0;1)}_{i,i+1}X_{i+1,i}\Lambda_{i,i+1}^{(0;1)}$, none of the chiral cycles contain the same arrow twice due to the implicit contractions with Levi-Civita tensors.

Expanding these cycles in terms of the component arrows, every term in \eref{chiral_cycles_Y10} gives rise to $(m+1)!$ chiral cycles. Since there are $m$ of these terms, we conclude there are $m (m+1)!$ chiral cycles. For $m=1$ we obtain 2 chiral cycles, which are just the 2 terms in the superpotential. For $m=2$, there are 12 chiral cycles, 2 for each of the $J$- and $E$-terms associated to the 6 Fermis in the quiver. 
        
Since all $\Lambda_{i,i+1}^{(0;1)}$ have a single index, we will drop the superindex $(0;1)$ and instead write its $SU(m+1)$ index explicitly. With this, the Grassmann variables and the corresponding chiral cycles become:
        \begin{align}
            \theta_{i}^{p} \ : \ \Lambda_{0,1}^{p(0)}\cdots \Lambda^{p(i)}_{i,i+1}X_{i+1,i}\Lambda_{i,i+1}^{p(i+1)}\Lambda_{i+1,i+2}^{p(i+2)} \cdots \Lambda_{m-1,m}^{p(m)}X_{m,0} ~,
        \end{align}
where $1 \le i \le m$ and $p$ runs over the elements of the symmetric group $S_{m+1}$ of $m+1$ elements $\{0,\cdots , m\}$. Using them we can write down the Grassmann variables associated to every chiral field, which are given by:
\beq
\begin{array}{cl}
            \Theta (X_{m,0})   &= \prod_{i=1}^{m}\prod_{p\in S_{m+1}}\theta_{i}^{p} \\[.15cm]
            \Theta (X_{i+1,i}) &= \prod_{p\in S_{m+1}}\theta_{i}^{p} \\[.15cm]
            \Theta (\Lambda_{i,i+1}^{\mu})  &= \left( \prod_{j=0}^{i}\prod_{p\in S_{m+1} ; p(i)=\mu}\theta_{j}^{p}\right) \left(  \prod_{j=i}^{m}\prod_{p\in S_{m+1} ; p(i+1)=\mu}\theta_{j}^{p} \right)
\end{array}
\eeq

We also need the intersection numbers between chiral fields and the fundamental cycles of the torus. We can choose the fundamental cycles such that they are:
        \begin{align}
            \ev{ \Lambda_{0,1}^{\mu},\gamma_{\alpha}} &= \delta_{\mu,\alpha}\nonumber \\[.15cm]
            \ev{X_{1,0},\gamma_{\alpha}}          &= 1 
        \end{align}
          
We are ready to evaluate the integral to obtain the Newton polynomial of the moduli space and identify the perfect matchings. We first note that:
\beq
\Theta (X_{m,0})\Theta (X_{i+1,i}) =  \Theta (\Lambda_{i,i+1}^{\mu})\Theta (X_{m,0}) = 0 ~.
\eeq
This is trivially true since $\Theta(X_{m,0})$ contains all the Grassmann variables. Similarly,
\beq
\Theta (X_{j+1,j})\Theta (\Lambda_{i,i+1}^{\mu}) = 0 ~,
\eeq
since $\Theta (X_{j+1,j})$ and $\Theta (\Lambda_{i,i+1}^{\mu})$ have a common factor $\prod_{p\in S_{m+1} ; p(i)=\mu}\theta_{j}^{p}$ for $j \le i$ and $\prod_{p\in S_{m+1} ; p(i+1)=\mu}\theta_{j}^{p}$ for $j \ge i$. Also for $i < j$, $\Theta (\Lambda_{i,i+1}^{\mu})$ and $\Theta (\Lambda_{j,j+1}^{\nu})$ have a common factor:
\beq
            \prod_{p\in S_{m+1} ; p(i)=\mu , p(j) = \mu}\theta_{j}^{p} ~.
\eeq
Note that since $p$ is invertible, the common factor is non-trivial if and only if $\mu \ne \nu$. Hence, for $\mu \ne \nu$,
\beq
\Theta (\Lambda_{i,i+1}^{\mu})\Theta (\Lambda_{j,j+1}^{\nu}) = 0 ~.
\eeq

With these results in mind, the only surviving integrals are:
\beq
\begin{array}{l}
            \int \prod_{i=1}^{m}\prod_{p\in S_{m+1}}\dd\theta_{i}^{p}\,\Theta (X_{m,0}) = 1 \\[.25cm]
            \int \prod_{i=1}^{m}\prod_{p\in S_{m+1}}\dd\theta_{i}^{p}\,\prod_{i=1}^{m}\Theta (X_{i,i-1})  = 1 \\[.25cm]
            \int \prod_{i=1}^{m}\prod_{p\in S_{m+1}}\dd\theta_{i}^{p}\,\prod_{i=1}^{m}\Theta (\Lambda_{i-1,i}^{\mu}) = 1
\end{array}
\eeq
 
The Newton polynomial for $Y^{1,0}(\mathbb{P}^{m})$ therefore becomes
        \begin{align}
             P(x^{\mu}) = X_{m,0} + \sum_{\mu=0}^{m} \left[\prod_{i=1}^{m}\Lambda_{i-1,i}^{\mu}\right]x^{\mu} + \prod_{i=1}^{m}X_{i+1,i}\prod_{\nu=0}^{m}x^{\nu} \label{Y10_newton_polynomial} ~,
        \end{align}
from where we can read off the chiral field content of all perfect matchings and determine their position in the toric diagram.

\paragraph{Complete perfect matchings.}
        
As explained in \eref{section_full_fields_pms}, we can reconstruct the entire perfect matchings from their chiral field content. We summarize them in the table below.
\beq
            \renewcommand{\arraystretch}{1.3}
            \begin{array}{|c|c|c|}
                \hline
                \mbox{      Point      }  & \mbox{      Chirals      } & \mbox{      Additional fields      } \\
                \hline
                v_0 & X_{m,0}        & \bar{\Lambda}^{(m+1-k;m+1-k)}_{i+k,i} \\
                              &                & \bar{\Gamma}_{i+k,i}^{(m-k-1;m-k)} \\
                \hline
                  v_{\mu}, \mu=1\ldots m+1       & \Lambda^{(0;1;\mu)}_{i,i+1}     & \Lambda^{(k-1;k;\mu)}_{i,i+k} ~ , \, \bar{\Lambda}^{(m+1-k;m+1-k;\mu)}_{i+k,i} \\
                              &                                 & \ \ \ \bar{\Gamma}_{i+k,i}^{(m-k-1;m-k;\mu)}~ , \, \Gamma_{i,i+k}^{(k+1;k+1;\mu)} \ \ \\
                \hline
                v_{m+2}  & X_{i+1,i}      & \bar{\Lambda}^{(m+1-k;m+1-k)}_{i+k,i} \\
                              &                & \Gamma_{i,i+k}^{(k+1;k+1)} \\
                \hline
            \end{array}
            \label{pms_Y10}     
\eeq

\section{Orbifolds of $\mathbb{C}^{m+2}$}

\label{section_orbifolds_Cm+2}

Orbifolds of $\mathbb{C}^{m+2}$ constitute a large class of Calabi-Yau singularities, which are obtained by orbifolding a discrete subgroup $G$ of the $SU(m+2)$ isometry of flat space.  We will focus on abelian orbifolds, i.e. those for which the subgroup $G$ is abelian. The purpose of this section is twofold. First, we will initiate the study of general abelian orbifolds of $\mathbb{C}^{m+2}$. The literature contains some interesting classification of the corresponding toric diagrams up to relatively large $m$ \cite{Hanany:2010cx,Davey:2010px,Hanany:2010ne}, but there is no study of the associated $m$-graded quivers. In addition, we apply to these orbifolds the techniques we introduced for calculating perfect matchings. Combinatorial details of these computations are presented in Appendix \ref{section_orbifold_pms_appendix}.

An abelian subgroup of $SU(m+2)$ can be decomposed as:
\beq
G \cong \mathbb{Z}_{k_{1}} \times \mathbb{Z}_{k_{2}} \times \cdots \times \mathbb{Z}_{k_{m+1}} ~.
 \label{orbifold_group}
\eeq

\paragraph{Quiver and orbifold action.}
The quivers contain $\abs{G}$ nodes, which are indexed by elements of $G$. The action of $G$ on $\mathbb{C}^{m+2}$ is stipulated by specifying $m+1$ elements $g_{\alpha} \in G$, which are required to generate $G$.\footnote{At this point it is worth emphasizing some standard facts. First, we note that the decomposition of the orbifold group $G$ into cyclic groups as in \eref{orbifold_group} is not unique.  Moreover, there can be multiple different orbifolds for the same cyclic groups. Fully specifying the orbifold under consideration requires determining a set of generators. Finally, given an orbifold, the generators can be picked in different ways.} We define:
\beq
g_{m+2} = -\sum_{\alpha = 1}^{m+1}g_{\alpha} ~.
\label{generator_constraint}
\eeq
            
We now discuss the matter content of the quiver, starting with the chiral fields. Fields of higher degree follow an analogous discussion. There are $(m+2)\abs{G}$ chiral fields. We can think about each of them as arising from an element $g \in G$ and a chiral field in $\mathbb{C}^{m+2}$ theory as follows:
\beq               
(g,\Phi^{(0;\mu)}) \to \Phi^{(0;\mu)}_{g,g+g_{\mu}} ~,      
               \label{orbifold_chirals}
\eeq
with $\mu=1,\cdots,m+2$.
            
The chiral fields in the unorbifolded $\mathbb{C}^{m+2}$ quiver theory transform in the fundamental representation of $SU(m+2)$. An orbifold gauges a discrete subgroup $G$ of the global symmetry of its parent theory. This means that generically there is no non-abelian global symmetry left. It is for this reason that in \eref{orbifold_chirals} we explicitly wrote the index $\mu$ of the parent field and omitted the $1$ that indicates the fundamental representation. Similarly, the arrows of degree $k$ are given by
            \begin{align}
                 (g,\Phi^{(k;\mu_{1}\cdots \mu_{k+1})}) \to \Phi^{(k;\mu_{1}\cdots\mu_{k+1})}_{g,g+g_{\mu_{1}}+\cdots + g_{\mu_{k+1}}} ~.
                 \label{orbifolds_all_fields}
             \end{align} 
As before, we have written the $k+1$ indices of the corresponding antisymmetric representation of $SU(m+2)$ in which the degree $k$ arrows in $\mathbb{C}^{m+2}$ transform.

\paragraph{Periodic quiver and superpotential.}

The periodic quiver of an orbifold theory is obtained by enlarging the fundamental domain of the $\mathbb{C}^{m+2}$ quiver. This enlargement is described by $m+1$ linearly independent points $v_{\alpha}$ in the integer lattice. The $v_{\alpha}$ are defined up to $SL(m+1,\mathbb{Z})$ transformations, which preserve the underlying torus. It is always possible to use $SL(m+1,\mathbb{Z})$ to take the $v_{\alpha}$ to triangular form, i.e. such that:
\beq
(v_{\alpha})_{\beta} = 0 \ \ \ \  \mbox{for } \beta > \alpha ~.
\eeq
                
The orbifold group \eref{orbifold_group} and its action can be determined from such a triangular $v_{\alpha}$. Let $u_{\alpha}$ be the first non-zero integer point on the segment connecting the origin to $v_{\alpha}$. Then:
\beq
                v_{\alpha} = k_{\alpha}u_{\alpha} ~,      
\eeq
fixes the integers $k_{\alpha}$ in \eref{orbifold_group}.

Every integer point in the enlarged torus can be written as $h_{\alpha}u_{\alpha}$ with $h_{\alpha} \in \mathbb{Z}_{k_{\alpha}}$. So $h\equiv (h_{1},\cdots,h_{m+1})$ is an element of $G$ and labels the node of the periodic quiver that is located at this point. In particular this is true for the $m+1$ unit vectors:
\beq
(1,0,\cdots,0) \ \ , \ \  (0,1,0\cdots,0) \ \ , \ \   \cdots \ \ , \ \   (0,\cdots,0,1) ~.
\eeq
The elements $g_{\alpha} \in G$ labeling these $m+1$ points are the generators defining the orbifold action.

We have explained how to locate the nodes on the $\mathbb{T}^{m+1}$ torus. Connecting them with the fields in \eref{orbifolds_all_fields}, we complete the periodic quiver. The superpotential consists of the minimal plaquettes in it. They are all cubic and we can explicitly write the superpotential, which is given by: 
\beq
W = \sum_{g\in G}\sum_{i+j+k = m+2}\epsilon_{\mu_{1}\cdots\mu_{m+2}}\Phi^{(j-1;\mu_{1}\cdots\mu_{j})}_{g,g+\mathfrak{g}(\mu;j)}\Phi^{(k-1;\mu_{j+1}\cdots\mu_{j+k})}_{g+\mathfrak{g}(\mu;j), g+\mathfrak{g}(\mu;j+k)}\bar{\Phi}^{(m+1-j-k;\mu_{j+k+3}\cdots\mu_{m+2})}_{g+\mathfrak{g}(\mu;j+k),g} ~, \label{general_orbifold_potential}
\eeq
where we defined:
\beq
\mathfrak{g}(\mu,j) = \sum_{\alpha=1}^{j}g_{\mu_{\alpha}} ~.
\eeq

The previous discussion can be immediately translated into an algorithm for the construction of the $m$-dimer for general abelian orbifolds of $\mathbb{C}^{m+2}$, which corresponds to the appropriate stacking of $|G|$ copies of the $(m+2)$-permutohedron.

\paragraph{From the orbifold action to the periodic quiver.}

Having explained how a given enlargement of the fundamental domain of the periodic quiver translates into the orbifold action, we now discuss the inverse problem. As previously mentioned, the decomposition of the orbifold group into cyclic groups is not unique. The important point is the relation among the generators and we can choose a decomposition that simplifies them. For this purpose, we take $\mathbb{Z}_{k_{1}}$ to be the cyclic group generated by $g_{1}$ and hence $g_{1}$ to be $(1,0,\cdots,0)$. Next, we take $\mathbb{Z}_{k_{1}} \times \mathbb{Z}_{k_{2}}$ to be the group generated by $g_{1}$ and $g_{2}$ so that $g_{2} = (g_{2,1},1,0,\cdots,0)$. Continuing with this process, we can choose $\mathbb{Z}_{k_{1}}\times \cdots \times \mathbb{Z}_{k_{\alpha}}$ to be the group generated by $g_{1},\cdots,g_{\alpha}$, so that: 
\beq
g_{\alpha} = (g_{\alpha,1},g_{\alpha,2},\cdots,g_{\alpha,\alpha-1},1,0,\cdots,0) \ \ \ \mbox{for all }1 \le \alpha \le m+1 ~.
\label{triangular_g_alpha}
\eeq 
This presentation of $g_{\alpha}$ makes it clear how to enlarge the fundamental domain of the periodic quiver of $\mathbb{C}^{m+2}$ to construct the periodic quiver for the action $g_{\alpha}$. The vectors $v_{\alpha}$ which result in this action are
        \begin{align}
            v_{1} &= k_{1}(1,0,0,\cdots,0) \nonumber \\ 
            v_{2} &= k_{2}(-g_{2,1},1,0,\cdots,0) \nonumber \\
            v_{3} &= k_{3}(-g_{3,1}+g_{3,2}g_{2,1} , -g_{3,2} , 1 , 0 , \cdots ,0) \nonumber \\
            &\vdots \\
            v_{m+1} &= k_{m+1}\Big(-g_{m+1,1}+g_{m+1,2}g_{2,1}+\cdots+(-1)^{m}\prod_{i=0}^{m-1}g_{m+1-i,m-i},\cdots,-g_{m+1,m},1\Big) \nonumber 
            \label{fundamental_domain}
        \end{align}

\paragraph{Chiral cycles.}
     
The chiral cycles for a general abelian orbifold of $\mathbb{C}^{m+2}$ can be determined applying the prescription presented in \sref{section_chiral_cycles_general_m} to the superpotential \eref{general_orbifold_potential}. It is however clearer and conceptually simpler to directly orbifold the chiral cycles of $\mathbb{C}^{m+2}$. The chiral cycles of $\mathbb{C}^{m+2}$ are indexed by elements of $S_{m+1}$, so there are $(m+1)!$ of them. Explicitly, to every $p \in S_{m+1}$ we associate the chiral cycle:
\beq
        \Phi^{(0;p(1))}\Phi^{(0;p(2))}\cdots \Phi^{(0;p(m+1)} \Phi^{(0;m+2)} ~.
        \label{chiral_cycles_Cm+2}
\eeq

For the orbifolds under consideration, the quiver contains $(m+2)\abs{G}$ chiral fields, given in \eref{orbifold_chirals}. There are $(m+1)! \abs{G}$ chiral cycles, which can be found by directly orbifolding the chiral cycles \eref{chiral_cycles_Cm+2}. They are:
\beq
        \theta^{p}_{g}  \ : \ \Phi^{(0;p(1))}_{g,g+g_{p;1}}\Phi^{(0;p(2))}_{g+g_{p;2},g+g_{p;2}}\cdots \Phi^{(0;p(m+1))}_{g+g_{p;m},g+g_{p;m+1}} \Phi^{(0;m+2)}_{g+g_{p;n},g} ~,
\eeq
where the $\theta^{p}_{g}$ are the corresponding Grassmann variables and   
\beq
        g_{p;\alpha} = \sum_{\beta=1}^{\alpha}g_{p(\beta)} ~.
\eeq

It is now straightforward to construct the $\Theta$ functions:
    \begin{align}
        \Theta(\Phi^{(0;m+2)}_{g,g+g_{m+2}}) &= \prod_{p \in S_{m+1}}\theta^{p}_{g-g_{m+2}} \nonumber \\
        \Theta(\Phi^{(0;\alpha)}_{g,g+g_{\alpha}}) &= \prod_{p\in S_{m+1}} \theta^{p}_{g-g_{p;p^{-1}(\alpha)}}
    \end{align}
    The Grassmann integral that generates the perfect matchings is:
\beq
        P(X_{i}) = \int \prod_{g\in G}\prod_{p \in S_{m+1}}\dd \theta^{p}_{g}\,\exp (\sum_{\mu=1}^{m+2}\sum_{g\in G}\Theta(\Phi^{(0;\mu)}_{g,g+g_{\mu}})\Phi^{(0;\mu)}_{g,g+g_{\mu}}\prod_{\alpha=1}^{m+1}x_{\alpha}^{\ev{\Phi^{\mu}_{g,g+g_{\mu}},\gamma_{\alpha}}}) ~. 
        \label{onp}
\eeq

\subsection{Orbifolds of $\mathbb{C}^{m+2}$ with $SU(m+2)$ global symmetry}

Orbifolding generically breaks the $SU(m+2)$ global symmetry of the parent $\mathbb{C}^{m+2}$ theory. In this section we consider the $\mathbb{Z}_{m+2}$ orbifolds with $g_{\alpha} = 1$ for all $\alpha$, which preserve the full $SU(m+2)$.\footnote{Since there is only one $\mathbb{Z}_{m+2}$ factor, we provide a single component for the $g_\alpha$'s.}

The toric diagram for these orbifolds consists of the following $m+3$ points:
            \begin{align}
                v_0=(0,\ldots,0)~, 
                \qquad 
                \begin{array}{l}
                v_1= (1,0,0,\ldots,0)~, \\
                 v_2=(0,1,0,\ldots,0)~, \\
                \quad \vdots \\
                 v_{m+1}=(0,0,\ldots,0, 1)~, 
                \end{array} 
                \qquad 
                v_{m+2}=(-1,-1, \ldots, -1)~.
                 \label{eq_toric_Cn_Zn}
            \end{align}

\paragraph{Quiver and superpotential.}

The corresponding quiver theories were first presented in \cite{Closset:2018axq} (for earlier appearances of the quivers for $m=3$ and $m=4$, see \cite{Diaconescu:2000ec,Douglas:2002fr,Franco:2016tcm,Closset:2017yte} and \cite{Franco:2017lpa}, respectively). They contain $m+2$ nodes and the following arrows:
\beq
                \Phi_{i,i+k}^{(k-1;k)} : i \xrightarrow[(k-1)]{\,\,\,\,\,\,\,\, \binom{m+2}{k}\,\,\,\,\,\,\,\,} i+k \qquad 0 \le i < m+2 ; 1 \le k < m+2 - i
\eeq
where $\Phi_{i,i+k}^{(k-1;k)}$ transforms in the $k$-index antisymmetric representation of $SU(m+2)$.

The superpotential is:
\beq
                W = \sum_{i+j+k < m+2}\Phi_{i,i+j}^{(j-1;j)}\Phi_{i+ji+j+k}^{(k-1;k)}\bar{\Phi}_{i+j+ki}^{(m+1-j-k;m+2-j-k)} ~ ,
                \label{potential_Cn_Zn}
\eeq
where we have suppressed the $SU(m+2)$ indices and used the convention in \eref{convention_products} for products.

\paragraph{Perfect matchings and moduli space.} 

Next let us compute the moduli spaces of these theories via the Grassmann integral \eqref{onp}. Appendix \ref{section_orbifold_pms_appendix} discusses the evaluation of this integral and the combinatorics of general orbifolds. In this case, we can choose $\gamma_{\mu}$ such that the non-zero intersection numbers are:
\begin{align}
\ev{\Phi^{(0;m+2)}_{0,1},\gamma_{\mu}} &= -1 \nonumber \\[.15cm]
\ev{\Phi^{(0;\alpha)}_{0,1},\gamma_{\mu}}    &= \delta_{\alpha,\mu}
\end{align}

The resulting Newton polynomial is
            \begin{multline}
                P(x_{\alpha}) = \prod_{\mu=1}^{m+2}\bar{\Phi}^{(0;1;\mu)}_{m+1,0} + \sum_{i=0}^{m}\prod_{\mu=1}^{m+2}\Phi^{(0;1;\mu)}_{i,i+1} + \sum_{\alpha=1}^{m+1}\left(\bar{\Phi}^{(0;1;\alpha)}_{m+1,0}\prod_{i=0}^{m}\Phi^{(0;1;\alpha)}_{i,i+1}\right) x_{\alpha} \\
                 + \left(\bar{\Phi}_{m+1,0}^{(0;1;m+2)}\prod_{i=0}^{m}\Phi_{i,i+1}^{(0;1;m+2)}\right)\prod_{\alpha=1}^{m+1}x_{\alpha}^{-1} \label{cn_zn_newton_polynomial}
            \end{multline}
The corresponding toric diagram is \eref{eq_toric_Cn_Zn}, as expected. 
           
There are $m+2$ perfect matchings at the internal point of the toric diagram $v_0$. Their chiral content is given by the first two terms in \eref{cn_zn_newton_polynomial}. Completing these perfect matchings, we obtain:
            \begin{align}
            \renewcommand{\arraystretch}{1.3}
            \begin{array}{|c|c|c|}
                \hline
                \ \mbox{        Perfect matching       }  \ \ & \mbox{      Chirals      } & \ \mbox{      Additional fields      } \ \ \\
                \hline
                s_{0}                         & \bar{\Phi}_{m+1,0}^{(0;1)} & \bar{\Phi}_{k,j}^{(m+1-k+j;m+2-k+j)} \,\,\,(k > j)   \\
                \hline
                s_{i} \,\,\,(1 \le i \le m+1) & \Phi_{i-1,i}^{(0;1)}       & \Phi_{jk}^{(k-j-1;k-j)} \,\,\,  (j < i \mbox{ and } j < k)  \\
                                              &                            & \bar{\Phi}^{(m+1-k+j;m+2-k+j)}_{k,j} \,\,\, (k > j \ge i)\\
                \hline
            \end{array}        
        \end{align} 
        $v_0$ is the only point invariant under $SU(m+2)$. This is nicely reflected by the perfect matchings above, all of which consist of full $SU(m+2)$ representations.

On the other hand, the corners break $SU(m+2)$ down to $SU(m+1)$ by singling out a direction. This fact is already reflected at the level of chiral fields of the corresponding perfect matchings, as can be seen in \eref{cn_zn_newton_polynomial}. This pattern of symmetry breaking greatly facilitates the completion of these perfect matchings, which proceeds analogously to the case of $\mathbb{C}^{m+2}$. After including the fields of higher degree the corner perfect matchings are:
        \begin{align}
            \renewcommand{\arraystretch}{1.3}
            \begin{array}{|c|c|c|}
                \hline
                \ \mbox{        Point in Toric Diagram       }  \ \ & \mbox{      Chirals      } & \ \mbox{      Additional fields      } \ \ \\
                \hline
                v_{\mu}                       & \Phi_{i,i+1}^{(0;1;\mu)}       & \Phi_{i,i+k}^{(k-1;k;\mu)}  \\
                                              & \bar{\Phi}_{m+1,0}^{(0;1;\mu)} & \bar{\Phi}_{i+ki}^{(m+1-k;m+2-k;\mu)}   \\
                \hline
            \end{array}        
        \end{align}    

These perfect matchings for this orbifolds were derived in \cite{Closset:2018axq} by global symmetry arguments instead of direct computation as in this section.

\section{Conclusions}

 \label{label_section_conclusions}
 
The open string sector of the topological B-model model on CY $(m+2)$-folds is described by $m$-graded quivers with superpotentials. This correspondence extends to general $m$ the well known connection between CY $(m+2)$-folds and gauge theories on the worldvolume of D$(5-2m)$-branes for $m=0,\ldots, 3$. The determination of the quiver theory associated to a given geometry and the inverse problem are, in practice, computationally challenging. In this paper we developed new powerful tools to tackle this problem.

We introduced $m$-dimers, which fully encode the $m$-graded quivers and their superpotentials, in the case in which the CY $(m+2)$-folds are toric. The basic ideas of this correspondence were previously outlined in \cite{Futaki:2014mpa,Franco:2016qxh,Franco:2017lpa,Closset:2018axq}. Remarkably, as it has been extensively studied for $m=1,2$, $m$-dimers significantly simplify the connection between geometry and $m$-graded quivers. A key result of this paper is the generalization of the concept of perfect matching, which plays a central role in this map, to arbitrary $m$. We provided two alternative definitions of perfect matchings, which are based on the superpotential \sref{section_generalized_pms} and on chiral cycles \sref{section_pm_chiral_cycles_general}. 

We studied the $m$-dimers associated $\mathbb{C}^{m+2}$, which are elegantly given by $(m+2)$-permutohedron bricks. The dimers for any other toric CY$_{m+2}$ can be constructed from orbifolds of $\mathbb{C}^{m+2}$, which are simply given by stacking multiple permutohedra, via partial resolution. We can thus regard this class of dimers as a universal parent theory in any dimension.

We also introduced various simplified methods for computing perfect matchings and the corresponding toric diagrams, culminating in the Grassmann integral given in \eref{general_Grassmann_integral}. This algorithm considerably supersedes the direct application of the perfect matching definition and provides a generalization of the Kasteleyn matrix approach to arbitrary $m$. In order to illustrate these ideas, we applied them to the $F_0^{(m)}$ and $Y^{1,0}(\mathbb{P}^m)$ infinite families of singularities and to abelian orbifolds of $\mathbb{C}^{m+2}$. In all these cases, we obtained new results about the perfect matchings, which provide a more complete picture of the map between quivers and geometry.

Exploiting these tools, we derived novel combinatorial results for singularities at arbitrary $m$. For the $F_0^{(m)}$ family, we showed that the number of perfect matchings is related to Dedekind numbers. For CY 3-folds, the behavior of perfect matching multiplicities under Seiberg dualities connecting different toric phases is controlled by cluster transformations. It is tempting to conjecture that a generalization of cluster algebras \cite{ClustI} based on the mutations of $m$-graded quivers exist. If so, the combinatorics of perfect matchings might provide a useful handle for elucidating them. 

Finally, we initiated a general study of the quiver theories for abelian orbifolds of $\mathbb{C}^{m+2}$, introducing methods for connecting the orbifold action to the periodic identification of the enlarged fundamental domain.

There are various interesting directions for further research. Here we mention a couple of them. It would be interesting to develop the general $m$ analogues of other central concepts in the study of dimers. Zig-zag paths are a prime example. For CY 3-folds, they play a fundamental role for mirror symmetry \cite{Feng:2005gw} and for the corresponding cluster integrable systems \cite{MR3185352,Eager:2011dp}. For $m=1,2$, zig-zag paths are given by the difference between perfect matchings associated to corners of the toric diagram \cite{GarciaEtxebarria:2006aq,Franco:2015tya}. It is thus natural to expect that our definition of perfect matchings will shed light on this problem.

Perfect matchings on dimers also appear in the context of melting crystal models of CY 3-folds \cite{Okounkov:2003sp,Iqbal:2003ds,Ooguri:2008yb,Ooguri:2009ri}. It would be interesting to study similar melting models for higher dimensional CY singularities and to investigate their physical interpretations.

\acknowledgments
 
We would like to C. Closset, E. Garc\'ia-Valdecasas, J. Guo, G. Musiker and Xingyang Yu for enjoyable discussions and related collaborations. The work of AH and SF was supported by the U.S. National Science Foundation grant PHY-1820721. AH was also supported by the U.S. National Science Foundation grant PHY-1519449.

\appendix

\section{Perfect matchings for $F_0^{(m)}$}

\label{section_pms_F0}

In \sref{section_F0m}, we defined all of the perfect matchings for the central point of the toric diagram of $F_0^{(m)}$ in terms of Boolean functions as follows. Given a monotonically increasing Boolean function $f$, we define a collection of fields $\tilde{p}_{f}$ as follows:
\beq
            \tilde{p}_{f} = \{X_{ij}|f(i) =0 \mbox{ and } f(j) = 1\} ~. \label{def_tilde_p_f}
\eeq
        Using $\tilde{p}_{f}$ we define:
\beq
            p_{f} = \tilde{p}_{f} \cup \{\bar{X}_{ji}|X_{ij} \notin \tilde{p}_{f}\} ~. \label{def_p_f}
\eeq
Let us now show that $p_{f}$ is indeed a perfect matching. By its definition, for every arrow $p_f$ contains either the arrow or its conjugate. Given the superpotential \eref{W_F0}, in order to show that it is a perfect matching we need to verify that for every $k \succ j \succ i$, $p_{f}$ contains exactly one of $X_{ij}$, $X_{jk}$ or $\bar{X}_{ki}$. We proceed as follows:
\begin{enumerate}
\item Let us suppose that $X_{ij} \in p_{f}$. Then $f(j) = 1$, which implies that $X_{jk} \notin p_{f}$. Also $f(i) = 0$ and since $k \succ j$ and $f$ is monotonic then $f(k) = 1$. Therefore $X_{ik} \in p_{f}$, which means that $\bar{X}_{ki} \notin p_{f}$. 

\item Next we consider the case when $X_{jk} \in p_f$. This means that $f(j) = 0$ and $f(k) = 1$ and since $j \succ i$ then $f(i) = 0$. Then, $X_{ij} \notin p_{f}$. Also $X_{ik} \in p_f$, which means that $\bar{X}_{ki} \notin p_{f}$.

\item Finally, we consider the case when neither $X_{ij}$ nor $X_{jk}$ are in $p_{f}$. Here we further divide the problem into two subcases:
                \begin{enumerate}
                      \item
                          $f(k) = 1$. Since $X_{jk} \notin p_{f}$ then $f(j) = 1$, which in turn means that $f(i) = 1$ since $X_{ij}\notin p_{f}$. Hence, $X_{ik}\notin p_{f}$, which means that $\bar{X}_{ki} \in p_{f}$.
                       \item
                          $f(k) = 0$. Since $k \succ j \succ i$, monotonicity means that $f(i) = f(j) = 0$. Hence  $X_{ik} \notin p_{f}$, which means that $\bar{X}_{ki} \in p_{f}$. 
                \end{enumerate}  
\end{enumerate}
This completes the proof of our assertion that $p_{f}$ is a perfect matching.

Next we show that if $f$ is not a constant function and $p_{f} = p_{g}$, then $f = g$. For this we first note that for any non-constant monotonic function $f(0,\cdots,0) = 0$ and $f(1,\cdots,1) = 1$. If $f$ and $g$ are distinct they differ at some argument $i$. Without loss of generality, we assume that $f(i) = 0$ and $g(i) = 1$. This means that $X_{i,(1,\cdots,1)} \in p_{f}$ but $X_{i,(1,\cdots,1)} \notin p_{g}$. Hence $p_{f}$ and $p_{g}$ are also distinct.

 Both the constant functions $f = 0$ and $f = 1$ are monotonic and for both of them $\tilde{p}_{f}$ is empty and hence they determine the same perfect matching, which is precisely the $p_0$ defined in \eref{p0}.
 
Going in the opposite direction, we want to prove that all the central perfect matchings for $F_{0}^{(m)}$ are determined by increasing Boolean functions. This can be achieved by assigning to every central perfect matching $p$ a monotonic Boolean function $f_{p}$ such that:
\beq
    p_{f_{p}} = p ~. \label{p_f_p_identity}
\eeq

 Let us define $f_{p}$ as follows:
\beq
    f_{p}(j) = 1 \, \Leftrightarrow \, \mbox{there exists an } i \mbox{ such that }  j \succ i \mbox{ and } X_{ij} \in p ~. \label{def_f_p}
\eeq
We start by showing that $f_{p}$ is monotonically increasing. Assuming $f_{p}(j)=1$ then there exists an $i$ such that $X_{ij} \in p$. For every $k \succ j$ there is a term in the superpotential:
\beq
    X_{ij}X_{jk}\bar{X}_{ki} ~.
\eeq
This means that $\bar{X}_{ki} \notin p$, which in turn means $X_{ik} \in p$. Hence $f_{p}(k) = 1$ and $f_{p}$ is monotonic.

To prove \eqref{p_f_p_identity} we need to show that if $X_{ij} \in p$ then $f_{p}(i) = 0$. Let us suppose that this is not the case and $f_{p}(i) = 1$. By the definition of $f_{p}$, there is some $l$ such that $X_{li} \in p$. Consider the term in the superpotential:
\beq
    X_{li}X_{ij}\bar{X}_{jl} ~.
\eeq
Since $X_{ij} \in p$, we must have $X_{li} \notin p$ which leads to a contradiction. Hence $f_{p}(i) = 0$. This completes our determination of the central perfect matchings.

\section{Perfect matchings for general orbifolds of $\mathbb{C}^{m+2}$}

\label{section_orbifold_pms_appendix}

In this appendix we explain how to evaluate the integral \eqref{onp} and discuss the resulting combinatorics. Since $\Theta(\Phi^{(0;\alpha)}_{g,g+g_{i}})$ is a product of $(m+1)!$ Grassmann variables, we only need the $\abs{G}^{th}$ power of the exponent in \eqref{onp} i.e. every perfect matching of an orbifold by $G$ has $\abs{G}$ chiral fields in it.

Given a collection $q$ of $\abs{G}$ chiral fields:
\beq
q  = \{ \Phi^{(0;\mu_{i})}_{g_{i},g_{i}+g_{\mu_{i}}} | 1\leq i \leq \abs{G}\} ~,
\eeq
$q$ represents a perfect matching if an only if:
        \beq
            \prod_{i=1}^{\abs{G}}\Theta(\Phi^{(0;\mu_{i})}_{g_{i},g_{i}2+g_{\mu_{i}}}) = \prod_{g\in G}\prod_{p \in S_{m+1}}\theta_{g}^{p} \label{orbifold_matching_criteria} ~.
        \eeq
        Below we present the implications of this condition for perfect matching on the $k$-dimensional ``faces'' of the toric diagram.

\subsection*{Corners}

Every chiral field in an orbifold descends from a chiral field $\Phi^{(0;\mu)}$ of the parent $\mathbb{C}^{m+2}$ theory. Each $\Phi^{(0;\mu)}$ gives rise to a perfect matchings corresponding to a corner of the toric diagram of $\mathbb{C}^{m+2}$. Analogously, every corner of the toric diagram of an orbifold is occupied by a single perfect matchings $q^\mu$ for which the chiral fields correspond to all the descendants of $\Phi^{(0;\mu)}$ i.e.
\beq
q^{\mu} = \{ \Phi^{(0;\mu)}_{g,g+g_{\mu}}| g\in G\} ~.
\eeq
It is straightforward to check that the $q^{\mu}$ satisfy \eqref{orbifold_matching_criteria} and that they are the only such collections containing the descendants of a single chiral field in $\mathbb{C}^{m+2}$. Therefore, there are $m+2$ corners, which is in agreement with the fact that the toric diagram for an orbifold of $\mathbb{C}^{m+2}$ is an $m+1$-dimensional simplex. The precise shape of this simplex, up to $SL(m+1,\mathbb{Z})$ transformations, is controlled by the specific orbifold action.

\subsection*{Edges}

Next we consider the perfect matchings that lie on the edge connecting the points corresponding to $q^{\mu}$ and $q^{\nu}$. The internal points on this edge mix the descendants of $\Phi^{(0;\mu)}$ and $\Phi^{(0;\nu)}$. 

Such perfect matchings admit an elegant description in terms of the quotient group $G/G_{\mu\nu}$ where $G_{\mu\nu}$ is the group generated by: 
\beq
                \set{g_{\rho}|\rho \ne \mu,\nu} ~.
\eeq
            The elements of $G/G_{\mu\nu}$ are cosets $[g]$ of $G_{\mu\nu}$ in $G$, i.e.
\beq
                [g] = \set{g + h | h \in G_{\mu\nu}} ~.      
\eeq 

Applying \eqref{orbifold_matching_criteria} to a perfect matching $p$ on this edge results in the condition that if $\Phi^{(0;\mu)}_{g,g+g_{\mu}} \in p$ then for all $h \in G_{\mu\nu}$ we must have $\Phi^{(0;\mu)}_{g+h,g+h+g_{\mu}} \in p$. Similarly, if $\Phi^{(0;\nu)}_{g,g+g_{\nu}} \in p$ then we must have $\Phi^{(0;\nu)}_{g+h,g+h+g_{\nu}} \in p$. Then, the chiral fields in $p$ can be organized in terms of cosets of $G_{\mu\nu}$, i.e. elements of the quotient group $G/G_{\mu\nu}$. Concretely, if we define a new set of fields $\mathcal{X}^{\rho}_{[g],[g]+[g_{\rho}]}$ as: 
\beq
\mathcal{X}^{\rho}_{[g],[g]+[g_{\rho}]} = \{\Phi^{(0;\rho)}_{g+h,g+h+g_{\rho}} | h \in G_{\mu\nu} \} ~,
\label{orbifold_variable_change}
\eeq
every perfect matching on this edge is a collection of $\mathcal{X}^{\rho}_{[g],[g]+[g_{\rho}]}$ with $\rho \in \{\mu,\nu\}$. Since all $\mathcal{X}^{\rho}_{[g],[g]+[g_{\rho}]}$ contain $\abs{G_{\mu\nu}}$ chiral fields, their number in a perfect matching must be $\abs{G}/\abs{G_{\mu\nu}} = \abs{G/G_{\mu\nu}}$. Every perfect matching of this edge can be written as:
\beq
q = \{ \mathcal{X}^{\rho}_{[g_{i}],[g_{i}]+[g_{\rho}]} | 1 \leq i \leq  \abs{G/G_{\mu\nu}} \}  ~.
\label{edge_matching_candidate}
\eeq
In order to determine whether such $q$ results in a perfect matching, we first note that $G/G_{\mu\nu}$ is a cyclic group generated by $[g_{\mu}] = -[g_{\nu}]$. Hence, $[g_{\mu}]$ defines the action of a $\mathbb{C}^{2}/(G/G_{\mu\nu})$ orbifold. We can regard $\mathcal{X}^{\mu}_{[g],[g]+[g_{\mu}]}$ and $\mathcal{X}^{\nu}_{[g],[g]+[g_{\nu}]}$ as the chiral fields of this orbifold.\footnote{Since $m=0$ for this orbifold, there is a subtlety resulting from the fact that in this case the conjugate of a chiral field also has degree 0. With this definition, the conjugate of $\mathcal{X}^{\mu}_{[g],[g]+[g_{\mu}]}$ is $\mathcal{X}^{\nu}_{[g]-[g_{\nu}],[g]}$}. Then \eqref{orbifold_matching_criteria} implies that $q$ is a perfect matching via the map given in \eqref{orbifold_variable_change} if and only if it is a perfect matching of the corresponding $\mathbb{C}^{2}/(G/G_{\mu\nu})$ orbifold. As already mentioned, there is no superpotential for $m=0$ theories, hence a perfecting matching simply corresponds to assigning an orientation to the unoriented chirals. There are $2^{\abs{G/G_{\mu\nu}}}$ of them, one for every subset of $\abs{G/G_{\mu\nu}}$. Given such a subset $s$, the corresponding perfect matching $q_{s}$ is:
\beq
q_{s} = \{ \mathcal{X}^{\mu}_{[g],[g]+[g_{\mu}]} | [g]\in s\}   \cup \{ \mathcal{X}^{\nu}_{[g]-[g_{\nu}],[g]} | [g]\notin s\} ~.
\eeq

\subsection*{Faces}

This behavior generalizes to faces of any dimension. There are $\binom{m+2}{k+1}$ faces of dimension $k$, one for each collection $\{\mu_{1},\cdots,\mu_{k+1}\}$ of $k+1$ coordinates of $\mathbb{C}^{m+2}$. The perfect matchings for such face only involve the descendants of $\Phi^{(0;\mu_{1})},\cdots,\Phi^{(0;\mu_{k+1})}$. These perfect matchings can be described in terms of the quotient group $G/G_{\mu_{1}\cdots\mu_{k+1}}$ where $G_{\mu_{1}\cdots\mu_{k+1}}$ is the subgroup of $G$ generated by:
\beq
\set{g_{\nu}|\nu \notin \{\mu_{1},\cdots,\mu_{k+1} \}} ~.     
\eeq
 $G/G_{\mu_{1}\cdots\mu_{k}}$ is useful because applying \eqref{orbifold_matching_criteria} to a perfect matching $q$ on this face results in the condition that if $\Phi^{(0;\mu)}_{g,g+g_{\mu}} \in q$ then for all $h \in G_{\mu_{1}\cdots\mu_{k+1}}$ we must also have $\Phi^{(0;\mu)}_{g+h,g+h+g_{\mu}} \in q$. These perfect matchings can hence be recast in terms of new fields defined using the cosets of $G_{\mu_{1}\cdots\mu_{k+1}}$ i.e. the elements of $G/G_{\mu_{1}\cdots\mu_{k+1}}$. These fields are given by:
\beq
\mathcal{X}^{\mu}_{[g],[g]+[g_{\mu}]} = \{ \Phi^{(0;\mu)}_{g+h,g+h+g_{\mu}} | h \in G_{\mu_{1}\cdots\mu_{k}} \} ~.
\label{quotient_variables}
\eeq
The condition stated above implies that the perfect matchings on this face can be written as a collection of $\mathcal{X}^{\mu}_{[g],[g]+[g_{\mu}]}$.
As in the case of edges, we can regard these new fields as the chiral fields of a $\mathbb{C}^{k+1}/(G/G_{\mu_{1}\cdots\mu_{k+1}})$ orbifold with action given by $[g_{\mu_{1}}], \cdots , [g_{\mu_{k}}]$. With this in mind, we can straightforwardly determine which collections of $\mathcal{X}^{\mu}_{[g],[g]+[g_{\mu}]}$ correspond to the perfect matchings. A set of $\mathcal{X}^{\mu}_{[g],[g]+[g_{\mu}]}$ is a perfect matching via the map in \eqref{quotient_variables} if and only if it is a perfect matching of the $\mathbb{C}^{k+1}/(G/G_{\mu_{1}\cdots\mu_{k+1}})$ orbifold described above.

The upshot of this discussion is that the perfect matchings on a $k$-dimensional face of an orbifold theory have a remarkably simple and natural description. This face is itself the toric diagram of an orbifold of $\mathbb{C}^{k+1}$. The multiplicities of perfect matchings on this face are the ones obtained from the quiver of the $\mathbb{C}^{k+1}$ orbifold. In addition, the perfect matchings themselves are related to the perfect matchings of this $\mathbb{C}^{k+1}$ orbifold by a simple map.

 \subsection*{Internal Points}

From the discussion presented above it is clear that a perfect matching at an internal point must contain at least one descendant of each $\Phi^{(0;\mu)}$ for $1 \le \mu \le m+2$. This innocuous statement is enough to rule out the existence of internal points for small orbifold groups. An internal point must contain at least $m+2$ chiral fields. On the other hand the number of chiral fields in a perfect matching of a $\mathbb{C}^{m+2}/G$ orbifold is $\abs{G}$. Hence the toric diagram has no internal points if $\abs{G} < m+2$.

\subsubsection*{An Example: $\mathbb{C}^{m+2}/\mathbb{Z}_{m+2}$ with $g_{\alpha} = 1$}

As an example of this discussion let us consider the $\mathbb{C}^{m+2}/\mathbb{Z}_{m+2}$ orbifold with all $g_{\alpha} = 1$, which was presented in detail in the main body of this work.

Since every $g_{\alpha}$ generates the orbifold group $G = \mathbb{Z}_{m+2}$, all the quotient groups $G/G_{\mu_{1}\cdots \mu_{k+1}}$ are trivial. Hence, every $k$-dimensional face is the toric diagram of $\mathbb{C}^{k+1}$ and has no points other than the corners. The perfect matchings at the corners are:
\beq
q^{\mu} = \{ \Phi^{(0;\mu)}_{i,i+1} | 1\leq i \leq m+2\} ~.     
\eeq
All that remains is to determine the perfect matchings at the internal point. Since the order of the orbifold group is $m+2$ any such perfect matching can be written as:
\beq
s = \{ \Phi^{(0;1)}_{i_{1},i_{1}+1},\Phi^{(0:2)}_{i_{2},i_{2}+1},\cdots , \Phi^{(0;m+2)}_{i_{m+2},i_{m+2}+1} \} ~,    
\eeq
which satisfies \eqref{orbifold_matching_criteria} if
\beq
i_{1} = i_{2} = \cdots = i_{m+2} = i ~.
\eeq
Hence we get one perfect matching for every element $i$ of $\mathbb{Z}_{m+2}$. The corresponding perfect matching $s_{i}$ is:
\beq
s_{i} = \{ \Phi_{i,i+1}^{(0;\mu)} | 1\leq \mu \leq m+2 \} ~.
\eeq
The right hand size can be recognized as the fundamental $SU(m+2)$ multiplet $\Phi_{i,i+1}^{(0;1)}$, which is to be expected from the $SU(m+2)$ invariance of the internal point.

\bibliographystyle{JHEP}
\bibliography{mybib}

\end{document}